\documentclass[twocolumn,floatfix,times,trackchanges]{aastex63}

\def\sun{\hbox{$\odot$}}
\def\earth{\hbox{$\oplus$}}
\def\lesssim{\mathrel{\hbox{\rlap{\hbox{%
 \lower4pt\hbox{$\sim$}}}\hbox{$<$}}}}
\def\gtrsim{\mathrel{\hbox{\rlap{\hbox{%
 \lower4pt\hbox{$\sim$}}}\hbox{$>$}}}}

\def\arcmin{\hbox{$^\prime$}}
\def\arcs{\hbox{$^{\prime\prime}$}}

\def\farcs{\hbox{$.\!\!^{\prime\prime}$}}

\def\micron{\hbox{$\mu$m}}

\newcommand{\Jyperbeam}{\mbox{~Jy~beam$^{-1}$}}
\newcommand{\mJyperbeam}{\mbox{~mJy~beam$^{-1}$}}

\newcommand{\kmpers}{\mbox{~km~s$^{-1}$}}

\newcommand{\Msun}{$M_\odot$}

\newcommand{\Tbol}{$T_\mathrm{bol}$}
\newcommand{\Mstar}{$M_\mathrm{\star}$}
\newcommand{\rstar}{$r_\mathrm{\star}$}
\newcommand{\Tstar}{$T_\mathrm{\star}$}
\newcommand{\Mdisk}{$M_\mathrm{disk}$}
\newcommand{\MDOTdisk}{$\dot{M}_\mathrm{disk}$}
\newcommand{\MDOTenv}{$\dot{M}_\mathrm{env}$}
\newcommand{\RdiskMAX}{$R_\mathrm{disk}$}

\newcommand{\Msunyr}{$M_\odot\,\mathrm{yr}^{-1}$}
\newcommand{\Rsun}{$R_\odot$}
\newcommand{\Rsub}{$R_\mathrm{sub}$}
\newcommand{\Rc}{$R_C$}

\newcommand{\acav}{$a_\mathrm{cav}$}
\newcommand{\bcav}{$b_\mathrm{cav}$}

\usepackage{CJKutf8}
\usepackage{graphicx}
\usepackage{longtable} \LTcapwidth=\textwidth
\usepackage{multirow}
\usepackage{subfigure}
\usepackage{ulem}
\usepackage{soul}
\usepackage{lipsum, babel}
\usepackage{enumitem}
\usepackage{amsmath}
\usepackage{euscript}
\usepackage{verbatim}
\usepackage{hyperref}

\setstcolor{red}

\submitjournal{ApJ}

\shorttitle{COM-rich Feature in G192.12-11.10}
\shortauthors{Hsu et al.}
\graphicspath{{./}{figures/}}

\begin{document}

\begin{CJK*}{UTF8}{bsmi}
\title{ALMASOP. A Rotating Feature Rich in Complex Organic Molecules in a Protostellar Core}

\author[0000-0002-1369-1563]{Shih-Ying Hsu}
\email{seansyhsu@gmail.com}
\affiliation{Institute of Astronomy and Astrophysics, Academia Sinica, No.1, Sec. 4, Roosevelt Rd, Taipei 106216, Taiwan (R.O.C.)}

\author[0000-0002-3024-5864]{Chin-Fei Lee}
\email{cflee@asiaa.sinica.edu.tw}
\affiliation{Institute of Astronomy and Astrophysics, Academia Sinica, No.1, Sec. 4, Roosevelt Rd, Taipei 106216, Taiwan (R.O.C.)}

\author[0000-0002-6773-459X]{Doug Johnstone}
\affiliation{NRC Herzberg Astronomy and Astrophysics, 5071 West Saanich Rd, Victoria, BC, V9E 2E7, Canada}
\affiliation{Department of Physics and Astronomy, University of Victoria, Victoria, BC, V8P 5C2, Canada}

\author[0000-0012-3245-1234]{Sheng-Yuan Liu}
\affiliation{Institute of Astronomy and Astrophysics, Academia Sinica, No.1, Sec. 4, Roosevelt Rd, Taipei 106216, Taiwan (R.O.C.)}

\author[0000-0002-5286-2564]{Tie Liu}
\affiliation{Key Laboratory for Research in Galaxies and Cosmology, Shanghai Astronomical Observatory, Chinese Academy of Sciences, 80 Nandan Road, Shanghai 200030, People’s Republic of China}

\author[0000-0002-9574-8454]{Leonardo Bronfman}
\affiliation{Departamento de Astronom\'{i}a, Universidad de Chile, Casilla 36-D, Santiago, Chile}

\author[0000-0002-9774-1846]{Huei-Ru Vivien Chen}
\affiliation{Department of Physics and Institute of Astronomy, National Tsing Hua University, Hsinchu, 30013, Taiwan}

\author[0000-0002-2338-4583]{Somnath Dutta}
\affiliation{Institute of Astronomy and Astrophysics, Academia Sinica, No.1, Sec. 4, Roosevelt Rd, Taipei 106216, Taiwan (R.O.C.)}

\author[0000-0002-5881-3229]{David J. Eden}
\affiliation{Department of Physics, University of Bath, Claverton Down, Bath BA2 7AY, UK}
\affiliation{Armagh Observatory and Planetarium, College Hill, Armagh BT61 9DB, UK}

\author[0000-0001-9304-7884]{Naomi Hirano}
\affiliation{Institute of Astronomy and Astrophysics, Academia Sinica, No.1, Sec. 4, Roosevelt Rd, Taipei 106216, Taiwan (R.O.C.)}

\author[0000-0002-5809-4834]{Mika Juvela}
\affiliation{Department of Physics, P.O.Box 64, FI-00014, University of Helsinki, Finland}

\author[0000-0003-2011-8172]{Kee-Tae Kim}
\affiliation{Korea Astronomy and Space Science Institute (KASI), 776 Daedeokdae-ro, Yuseong-gu, Daejeon 34055, Republic of Korea}
\affiliation{University of Science and Technology, Korea (UST), 217 Gajeong-ro, Yuseong-gu, Daejeon 34113, Republic of Korea}

\author[0000-0002-4336-0730]{Yi-Jehng Kuan}
\affiliation{Department of Earth Sciences, National Taiwan Normal University, Taipei, Taiwan (R.O.C.)}
\affiliation{Institute of Astronomy and Astrophysics, Academia Sinica, No.1, Sec. 4, Roosevelt Rd, Taipei 106216, Taiwan (R.O.C.)}

\author[0000-0003-4022-4132]{Woojin Kwon}
\affiliation{Department of Earth Science Education, Seoul National University, 1 Gwanak-ro, Gwanak-gu, Seoul 08826, Republic of Korea}
\affiliation{SNU Astronomy Research Center, Seoul National University, 1 Gwanak-ro, Gwanak-gu, Seoul 08826, Republic of Korea}
\affiliation{The Center for Educational Research, Seoul National University, 1 Gwanak-ro, Gwanak-gu, Seoul 08826, Republic of Korea}

\author[0000-0002-3179-6334]{Chang Won Lee}
\affiliation{Korea Astronomy and Space Science Institute (KASI), 776 Daedeokdae-ro, Yuseong-gu, Daejeon 34055, Republic of Korea}
\affiliation{University of Science and Technology, Korea (UST), 217 Gajeong-ro, Yuseong-gu, Daejeon 34113, Republic of Korea}

\author[0000-0003-3119-2087]{Jeong-Eun Lee}
\affiliation{Department of Physics and Astronomy, Seoul National University, 1 Gwanak-ro, Gwanak-gu, Seoul 08826, Korea}

\author[0000-0003-1275-5251]{Shanghuo Li}
\affiliation{School of Astronomy and Space Science, Nanjing University, 163 Xianlin Avenue, Nanjing 210023, People’s Republic of China}
\affiliation{Key Laboratory of Modern Astronomy and Astrophysics (Nanjing University), Ministry of Education, Nanjing 210023, People's Republic of China}

\author[0000-0002-6868-4483]{Sheng-Jun Lin}
\affiliation{Institute of Astronomy and Astrophysics, Academia Sinica, No.1, Sec. 4, Roosevelt Rd, Taipei 106216, Taiwan (R.O.C.)}

\author[0000-0002-1624-6545]{Chun-Fan Liu}
\affiliation{Institute of Astronomy and Astrophysics, Academia Sinica, No.1, Sec. 4, Roosevelt Rd, Taipei 106216, Taiwan (R.O.C.)}

\author[0000-0001-8315-4248]{Xunchuan Liu}
\affiliation{Shanghai Astronomical Observatory, Chinese Academy of Sciences, Shanghai 200030, PR China}

\author[0000-0002-5845-8722]{J. A. López-Vázquez}
\affiliation{Institute of Astronomy and Astrophysics, Academia Sinica, No.1, Sec. 4, Roosevelt Rd, Taipei 106216, Taiwan (R.O.C.)}

\author[0000-0003-4506-3171]{Qiuyi Luo}
\affiliation{Shanghai Astronomical Observatory, Chinese Academy  of Sciences, Shanghai 200030, People’s Republic of China}
\affiliation{School of Astronomy and Space Sciences, University of Chinese Academy of Sciences, No. 19A Yuquan Road, Beijing 100049, People’s Republic of China} 
\affiliation{Key Laboratory of Radio Astronomy and Technology, Chinese Academy of Sciences, A20 Datun Road, Chaoyang District, Beijing, 100101, P. R. China}

\author[0000-0002-6529-202X]{Mark G. Rawlings}
\affiliation{Gemini Observatory/NSF NOIRLab, 670 N. A’ohoku Place, Hilo, Hawai’i, 96720, USA}

\author[0000-0002-4393-3463]{Dipen Sahu}
\affiliation{Physical Research laboratory, Navrangpura, Ahmedabad, Gujarat 380009, India}
\affiliation{Institute of Astronomy and Astrophysics, Academia Sinica, No.1, Sec. 4, Roosevelt Rd, Taipei 106216, Taiwan (R.O.C.)}

\author[0000-0002-7125-7685]{Patricio Sanhueza}
\affiliation{Department of Astronomy, School of Science, The University of Tokyo, 7-3-1 Hongo, Bunkyo, Tokyo 113-0033, Japan}

\author[0000-0001-8385-9838]{Hsien Shang (尚賢)}
\affiliation{Institute of Astronomy and Astrophysics, Academia Sinica, No.1, Sec. 4, Roosevelt Rd, Taipei 106216, Taiwan (R.O.C.)}

\author[0000-0002-8149-8546]{Kenichi Tatematsu}
\affiliation{Nobeyama Radio Observatory, National Astronomical Observatory of Japan, National Institutes of Natural Sciences, 462-2 Nobeyama, Minamimaki, Minamisaku, Nagano 384-1305, Japan}
\affiliation{Department of Astronomical Science, The Graduate University for Advanced Studies, SOKENDAI, 2-21-1 Osawa, Mitaka, Tokyo 181-8588, Japan}

\author[0000-0001-8227-2816]{Yao-Lun Yang}
\affiliation{Star and Planet Formation Laboratory, RIKEN Cluster for Pioneering Research, Wako, Saitama 351-0198, Japan}



\begin{abstract} 
Interstellar complex organic molecules (COMs) in solar-like young stellar objects (YSOs), particularly within protostellar disks, are of significant interest due to their potential connection to prebiotic chemistry in emerging planetary systems.
We report the discovery of a rotating feature enriched in COMs, including CH$_3$OH, CH$_3$CHO, and NH$_2$CHO, in the protostellar core G192.12-11.10. 
By constructing a YSO model, we find that the COM-rich feature is likely located within or near the boundary of the Keplerian disk. 
The image synthesis results suggest that additional heating mechanisms leading to a warm ring or a warm inner disk are required to reproduce the observed emission. 
We discuss possible origins of the COM-rich feature, particularly accretion shocks as a plausible cause for a warm ring. 
Additionally, molecules such as C$^{18}$O, H$_2$CO, DCS, H$_2$S, and OCS exhibit distinct behavior compared to CH$_3$OH, indicating a range of physical and chemical conditions within the region. 
The observed kinematics of H$_2$S and OCS suggest that OCS resides in regions closer to the central protostar than H$_2$S, consistent with previous experimental studies.
\end{abstract}

\keywords{Protostars --- Complex organic molecules --- Star formation --- Circumstellar disks --- Interstellar medium --- Astrochemistry --- Pre-biotic astrochemistry}

\section{Introduction}
\label{sec:Intro}

Interstellar complex organic molecules (COMs) in solar-like young stellar objects (YSOs) are of great interest due to their potential connection to prebiotic chemistry in emerging planetary systems.
These COMs are thought to form on the icy mantles of dust grains via grain-surface reactions.
During the protostellar stage, they are released into the gas phase primarily through thermal desorption, triggered when the temperature rises above the ice sublimation threshold of approximately 100 K \citep[e.g.,][]{2009Herbst_COM_review,2020Jorgensen_review}.
In addition, several non-thermal desorption mechanisms have been proposed to explain the presence of gas-phase COMs in colder environments \citep[e.g.,][]{2010Oberg_B1b,2013Vasyunin_reactive_desorption,2016Jimenez-Serra_L1544,2021Wakelam_non-thermal_desorption}.

A major breakthrough in observational studies was the discovery of hot corinos, as named by \citet{2004Ceccarelli_HotCorino}, which are localized warm regions rich in COMs surrounding solar-like YSOs.
In the first decade following the discovery of hot corinos, only a small number of protostellar cores were found to host warm COMs.
Thanks to advances in technology and observatories such as the Atacama Large Millimeter/sub-millimeter Array (ALMA), recent survey-type studies have discovered a good number of protostellar cores harboring COMs \citep[e.g., ][]{2020Belloche_CALYPSO,2021Yang_PEACHES,2020Hsu_ALMASOP,2022Hsu_ALMASOP,2022Bouvier_ORANGES}. 
Moreover, \citet{2023Hsu_ALMASOP} demonstrated that gas-phase COMs can be ubiquitous in protostellar cores, provided that the innermost envelope is sufficiently heated by the central protostar. 
Other mechanisms in addition to the heating from the central protostar can also contribute to COM desorption, resulting in a variety of morphologies and kinematics for the observed COM emission.
To date, COMs have been detected in various YSO environments, including the innermost envelope \citep[e.g.,][]{2019Jacobsen_L483_COM}, outflow shock fronts \citep[e.g.,][]{2008Arce_L1157-B1_COMs}, jet bases \citep[e.g.,][]{2024Hsu_ALMASOP}, disk winds \citep{2024DeSimone_IRAS4A2_COMdiskwind}, and disk atmosphere \citep[e.g.,][]{2017Lee_HH212,2019Lee_HH212_COM_atm}.

Particularly, the presence of COMs near disks is of considerable importance, as these disks serve as the nurseries for planets, comets, and asteroids.
So far, only a handful of disks have reported the detection of COMs, such as those in 
a Class~0 protostellar core HH~212 \citep{2016Codella_HH212_H2O_COM,2017Lee_HH212,2019Lee_HH212_COM_atm,2022Lee_HH212_stratification}, 
a Class I/II protostellar core V883 Ori \citep{2018vantHoff_V883Ori_CH3OH,2019Lee_V883-Ori_outburst,2025Jeong_V883Ori_COM,2025Fadul_V883Ori}, 
a T Tauri star TW Hya \citep{2016Walsh_TWHya_CH3OH}, 
a T Tauri star  DG Tau \citep{2015Loomis_DMTau}, 
a Herbig Ae star HD 163296 \citep{2019Carney_HD163296_COM},
a young star Oph IRS 48 \citep{2021vanderMarel_OphIRS48_H2CO_CH3OH,2022Brunken_OphIRS48_COM} 
and an A-type star HD 100546 \citep{2021Booth_HD100546_COM}. 
Since the presence of gas-phase COMs requires a desorption mechanism, studying these molecules also provides valuable insights into the physical processes occurring within  disks.
For example, in HH~212, some COMs near the disk edge are proposed to be released by heat from accretion shocks, while others farther from the edge may be desorbed by radiation from the central protostar or interactions with disk winds \citep{2017Lee_HH212, 2017Tabone_HH212_cavity,2022Lee_HH212_stratification}.
In V883 Ori, COMs were freshly desorbed due to a stellar outburst \citep{2019Lee_V883-Ori_outburst,2025Jeong_V883Ori_COM}.
To advance our understanding of organic chemistry in protostellar disks, it is crucial to investigate more protostellar systems.

Here, we report the discovery of a COM-rich rotating feature likely residing in the disk of the protostellar core G192.12-11.10 (05$^\mathrm{h}$32$^\mathrm{m}$19$^\mathrm{s}$\!\!.368 +12$\degr$49$\arcmin$40$\arcs$\!\!.91), which is located in the $\lambda$-Orionis cloud \citep[distance $d=445\pm50$ pc;][]{2011Lombardi_2MASS_extinction_IV}.
This protostellar core was previously identified as a source of warm COM emission in the ALMA Survey of Orion PGCCs (ALMASOP) project \citep{2020Hsu_ALMASOP,2022Hsu_ALMASOP}. In this paper, we further reveal the rotating nature of the COM within the core.
We introduce the observational programs in \S \ref{sec:Obs} and present the observed morphologies and kinematics in \S \ref{sec:results}.
Based on our observations, we construct a model for the YSO, which is described in \S \ref{sec:model}.
In \S \ref{sec:Disc}, we discuss the possible origin of the COM-rich rotating feature and its implications.
Finally, we summarize our conclusions in \S \ref{sec:Conclusions}.


\begin{figure*}[htb!]
\centering
\includegraphics[width=\textwidth]{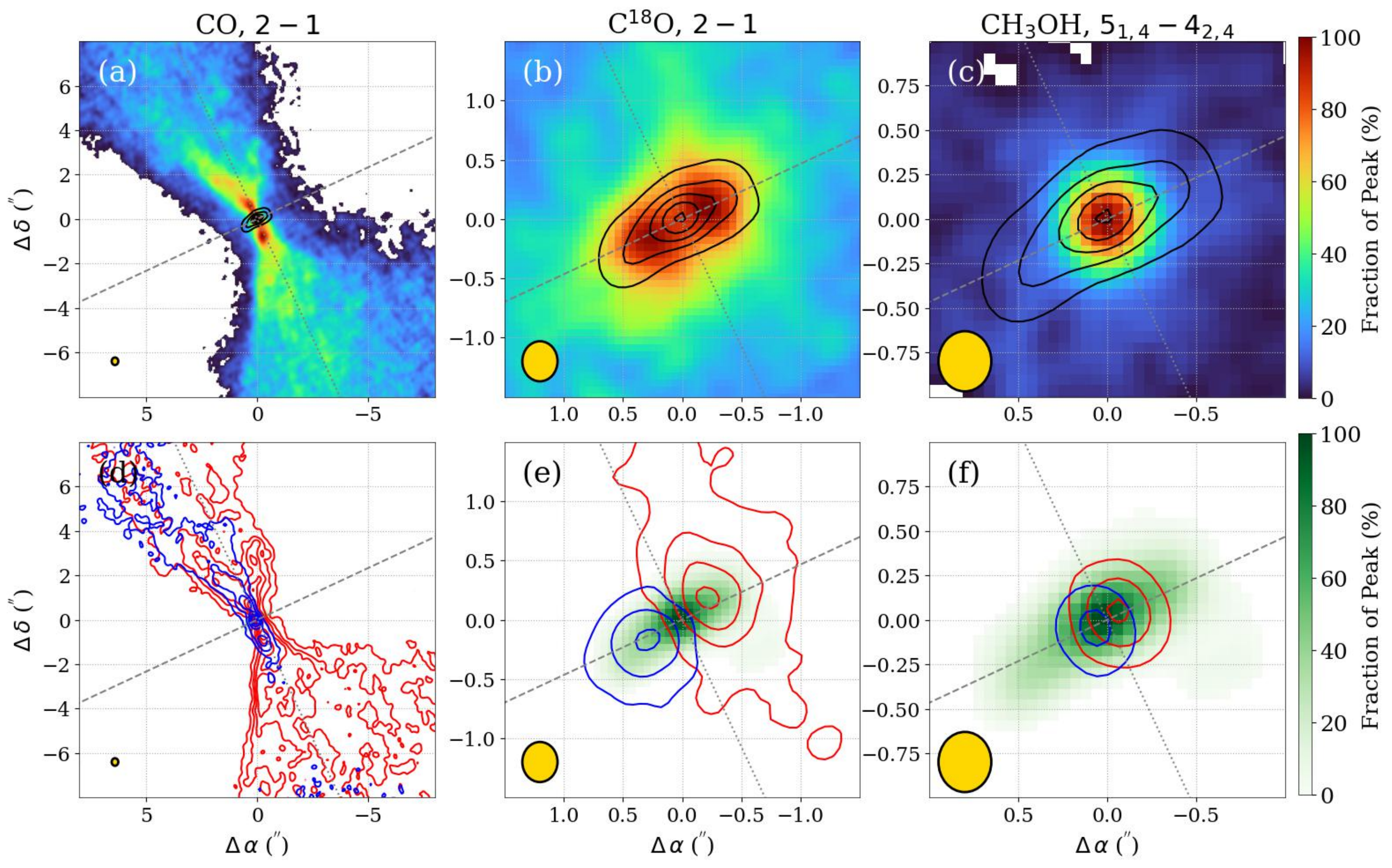}
\caption{\label{fig:img_main} 
The 1.3~mm continuum and the integrated intensity images for CO $J=2-1$, C$^{18}$O $J=2-1$, and CH$_3$OH $5_{1,4}$ -- $4_{2, 4}$ transitions. 
For each column, the label at the top shows the molecule species and the quantum number of the corresponding transition.  
In each panel, the dashed and dotted lines illustrate the axes across and along the outflow, respectively. 
\textit{Top}: Integrated intensity images (rainbow rasters) overlaid with 1.3~mm continuum (black contours). 
The color scales of the raster maps are shown as fractions of the peaks, which are 2.53, 0.56, and 0.86 \Jyperbeam \kmpers, respectively. 
The contour levels start from 20$\sigma$ with a step of 20$\sigma$. 
\textit{Bottom}: 1.3~mm continuum (green rasters) overlaid with integrated intensity images for blue- and red-shifted channels (blue and red contours, respectively). 
The color scales of the raster maps are shown as the fraction of the peak, which is 39 \mJyperbeam. 
The integration ranges for the CO, C$^{18}$O, and CH$_3$OH are 25.0, 4.05, and 7.50~\kmpers, respectively, with a center velocity at $v_\mathrm{LSR}=10$~\kmpers. 
The contour levels start from 10$\sigma$ with a step of 10$\sigma$. 
The gold eclipse in each panel represent the beam size around 0\farcs{42}. 
}
\end{figure*}

\section{Observations}
\label{sec:Obs}
This study mainly uses data obtained from the ALMA Cycle~6 program \#2018.1.00302.S, (PI: Tie Liu) in Band~6 (1.3~mm or 230~GHz) with the combination of both 12-m array (configurations C43–5 and C43–2) and 7-m array (also known as the Atacama Compact Array, ACA, or Morita Array). 
The projected baselines ranged from 6 to 965 k$\lambda$, and the resulting maximum recoverable scale was approximately 26\arcs.
The four spectral windows centered at 216.6, 218.9, 231.0, and 233.0 GHz have a uniform bandwidth of 1,875~MHz with a spectral resolution of 1.129~MHz (1.4~\kmpers\ at 230~GHz). 

We used \texttt{tclean} in CASA \citep[][]{casa:2022} with a robust value of 0.5. 
The resulting angular resolution was $\sim$ 0\farcs{42} \citep[corresponds to approximately 190 au for a distance of $\sim445$~ pc, ][]{2011Lombardi_2MASS_extinction_IV}. 
The sensitivities of the image reach $\sim$0.43~\mJyperbeam\ (0.10~K) and $\sim$3.5~\mJyperbeam\ (0.84~K) for the full--band continuum and each channel, respectively. 
For more details on the ALMASOP project, please refer to \citet{2020Dutta_ALMASOP}. 
%

\begin{figure}[htb!]
\centering
\includegraphics[width=.95\linewidth]{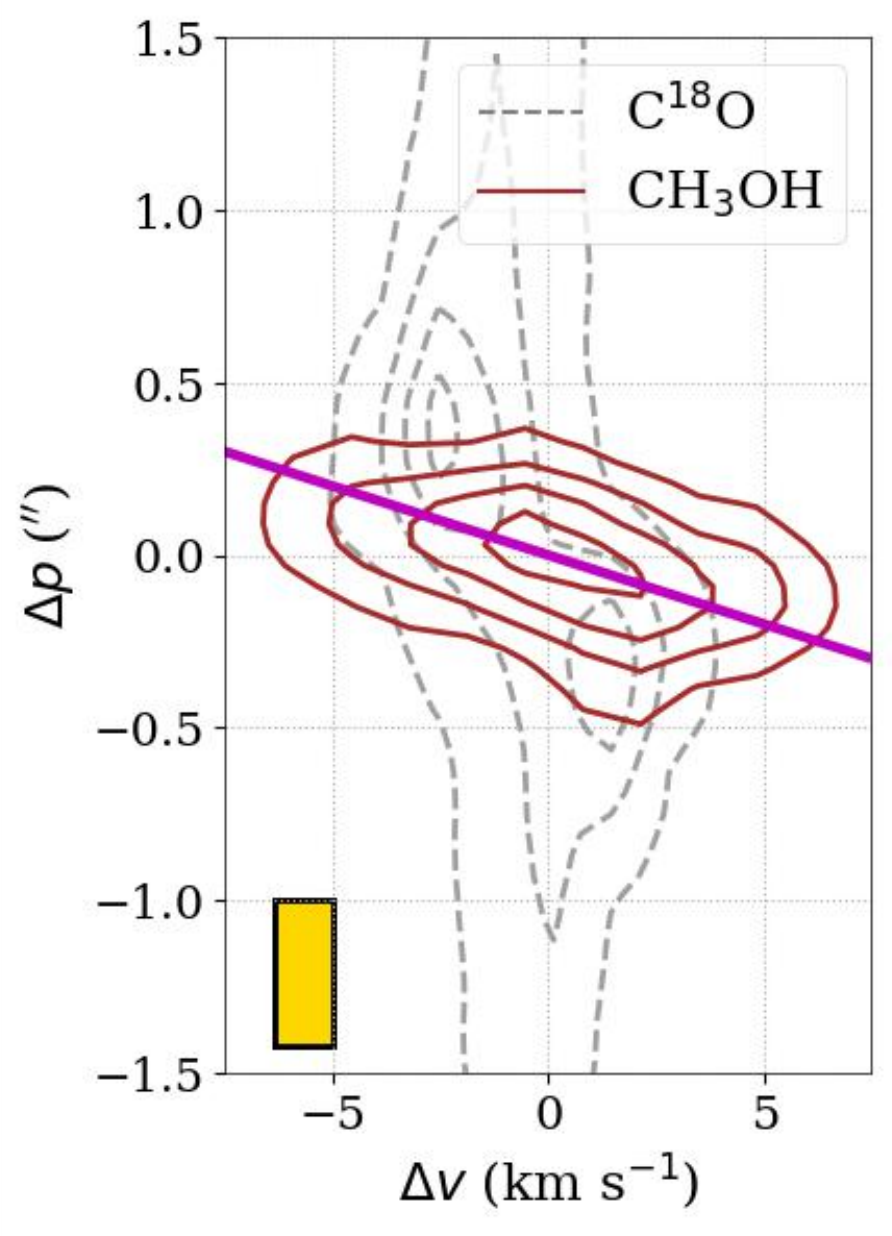}
\caption{\label{fig:PV_main} 
The PV diagrams for C$^{18}$O $J=2-1$ and CH$_3$OH $5_{1,4}$ -- $4_{2, 4}$ transitions. 
The diagrams were extracted by cutting along the continuum major axis from southeast (blue-shifted) to northwest (red-shifted) through the continuum peak. 
The flux density was averaged by three pixels (0\farcs{18}). 
The position reference for calculating the relative position ($\Delta p$) is set at the 1.3~mm continuum peak position. 
The velocity reference for calculating the relative velocity ($\Delta v$) is the local-standard-of-rest velocity at 10 \kmpers. 
The gold rectangles at the bottom left of each panel represent the beam size and spectral resolution.
In panel (a), the grey dashed contours and brown solid contours represent the C$^{18}$O and CH$_3$OH, respectively. 
The magenta line is the corresponding linear relation in Eq. \ref{eq:dp_dv}.  
}
\end{figure}

\section{A First Look at the Observed Features}
\label{sec:results}

\subsection{Morphology}
\label{sec:morphology}

Figure \ref{fig:img_main}(a) and (b) show the integrated intensity images of CO $J=2-1$ and C$^{18}$O $J=2-1$ transitions, respectively, overlaid with 1.3~mm dust continuum. 
The CO $J=2-1$ integrated intensity image reveals wide and extended outflow lobes having a position angle (PA) of $25^\circ$ (i.e., oriented along a northeast-southwest axis).
The presence of red-shifted outflow in both lobes suggests that the system is likely close to an almost edge-on system.
The dust continuum is distinctly elongated in this northwest-southeast direction, perpendicular to the outflow axis.
In Figure \ref{fig:img_main}(b), the C$^{18}$O $J=2-1$ emission is spatially extended, displaying two peaks aligned with the major axis of the dust continuum, each located on opposite sides of the outflow axis. 
The distance between the two peaks is roughly 0\farcs{5} (or $\sim220$ au). 
These morphologies suggest that the continuum is illustrating the inner dense envelope-disk system, and C$^{18}$O is predominately tracing the circumstellar envelope. 

Figure \ref{fig:img_main}(e) presents the integrated intensity images for the blue- and red-shifted channels separately.
The extended C$^{18}$O gas exhibits rotation with an axis along the outflow.
The western and eastern regions are red- and blue-shifted, respectively.
Moreover, the blue and red peaks in Figure \ref{fig:img_main}(e) coincide with the two peaks in Figure \ref{fig:img_main}(b). 
Consistently, as implied by Figure \ref{fig:img_main}(d), the molecular outflow likely rotates around an axis directed toward the northeast.
Additionally, in the northeast region, the significant overlap between the blue- and red-shifted outflows suggests that the system's geometry is predominantly edge-on.

Interestingly, as shown in Figures \ref{fig:img_main}(c) and (f), CH$_3$OH appears to be rotating along an axis consistent with the envelope.
Similar to C$^{18}$O, the blue- and red-shifted integrated intensity images of CH$_3$OH show distinct peaks; however, the separation between the blue and red peaks is significantly smaller than that of C$^{18}$O.
A notable difference is that the integrated intensity image of CH$_3$OH across the full spectrum, as shown in Figure \ref{fig:img_main}(c), does not exhibit two peaks or significant elongation but instead remains compact.
This suggests that CH$_3$OH gas is residing in the inner rotating regions near the central protostar, possibly the disk or the inner envelope. 


\begin{figure*}[htb!]
\centering
\includegraphics[width=.99\linewidth]{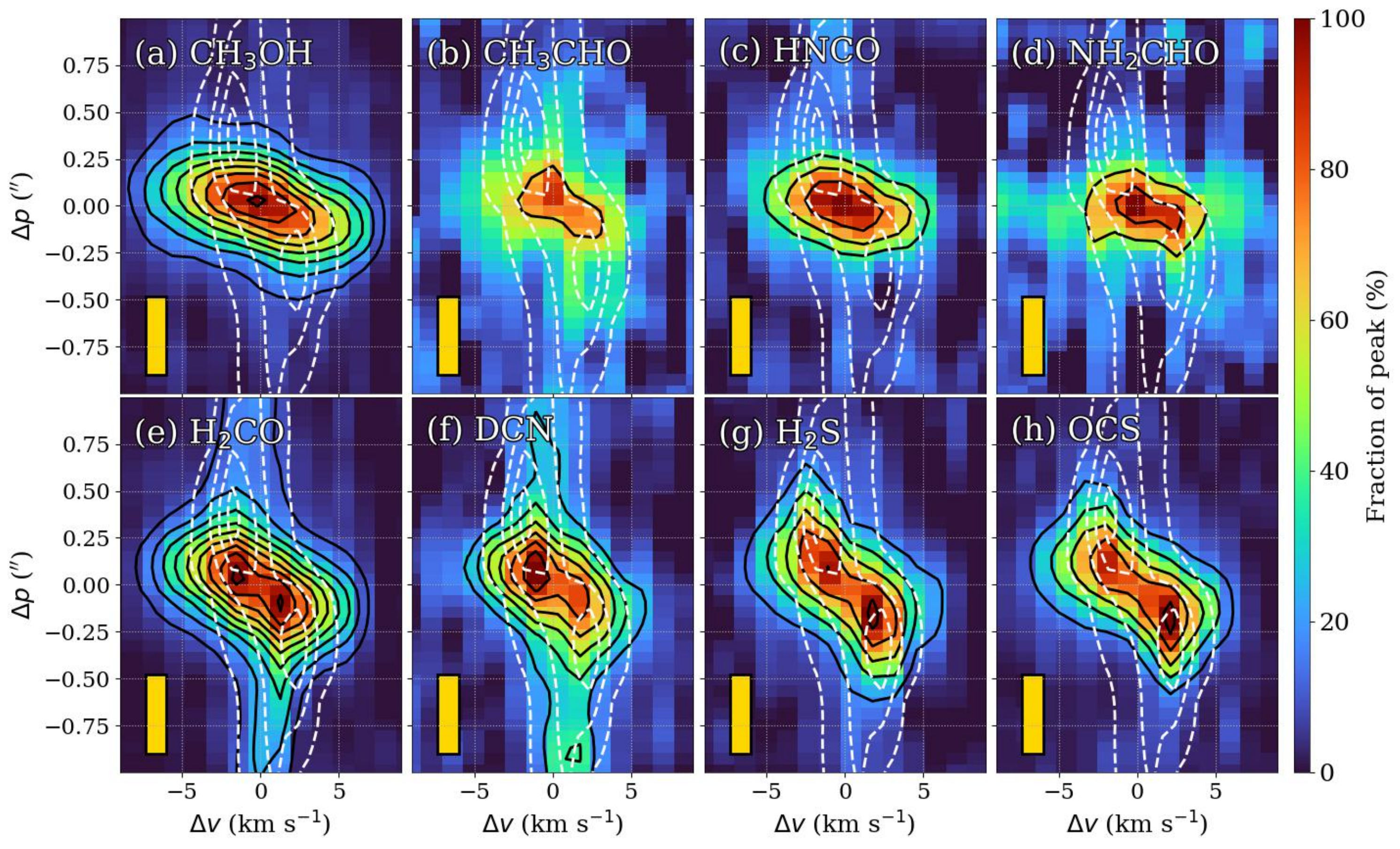}
\caption{\label{fig:PV_formulae_main} 
The observed PV diagrams of various molecules (represented by rasters and black solid contours) are overlaid with those of C$^{18}$O (shown as white dashed contours).
The PV diagrams were extracted with a path cutting across the outflow axis, as shown by the dashed line in Figure \ref{fig:img_main}, with values averaged over three pixels (0\farcs{18}.)
The rasters indicate the fractional intensity relative to their peak values.
For C$^{18}$O, the contour levels start at 5$\sigma$ and increase in steps of 5$\sigma$.
Similarly, for CH$_3$OH (a), H$_2$CO (e), H$_2$S (g), and OCS (h), the contour levels also begin at 5$\sigma$ with increments of 5$\sigma$.
For all other molecules, the contour levels start at 5$\sigma$ and increase in steps of 2.5$\sigma$.
The gold rectangles at the bottom left of each panel represent the beam size and spectral resolution.
}
\end{figure*}

\subsection{Kinematics}
\label{sec:kinematics}

To look at the kinematics of the envelope-disk system, we extracted position-velocity (PV) diagrams across the outflow axis (the major axis of the continuum, as illustrated by the dashed line in Figure \ref{fig:img_main}).
Figure~\ref{fig:PV_main} presents the PV diagrams for the C$^{18}$O $J=2-1$ and CH$_3$OH $5_{1,4}$ -- $4_{2,4}$ transitions.
Both diagrams show that the eastern region ($\Delta p > 0$) is blue-shifted, while the western region ($\Delta p < 0$) is red-shifted, indicating that C$^{18}$O and CH$_3$OH share the same rotation direction.
Despite this similarity, the PV diagrams of C$^{18}$O and CH$_3$OH exhibit distinct features, emphasizing their differing physical origins.

C$^{18}$O shows a typical pattern associated with an infalling and rotating envelope \citep[e.g.,][]{2014Sakai_L1527_envelope, 2014Oya_IRAS15398-3359, 2017Lee_HH212}. 
On the one hand, the PV diagram features an elongated distribution close to the region where $\Delta v=0$. 
On the other hand, taking the red-shifted part as an example, as the relative position approaches the center, $|\Delta v|$ gradually increases before dropping at a specific relative position.
This feature serves as an indicator of the disk-envelope interface, although its precise derivation depends on the specific model used. 

In contrast, the velocity gradient of CH$_3$OH appears to be linear. 
This linear pattern indicates that CH$_3$OH is exhibiting rotation within a ring.  
The maximum relative speed ($|\Delta v|$) of CH$_3$OH exceeds that of C$^{18}$O, supporting that the former is tracing the more inner region than the latter. 
Also, the CH$_3$OH PV diagram tentatively shows a high-velocity tail at $\Delta v < -5$~\kmpers, hinting at its possible presence in the disk region. 
The linearity can be described as: 
\begin{equation}
    \label{eq:dp_dv}
    \left ( \frac{\Delta p}{0.20~\mathrm{arcsec}} \right )=\left ( \frac{\Delta v}{5~\mathrm{km~s^{-1}}} \right ). 
\end{equation}
Assuming that the disk mass is much smaller than the stellar mass and an edge-on geometry, the relative velocity along the line of sight ($\Delta v$) at a relative position ($\Delta p$) following of a ring with Keplerian rotation is:
\begin{equation}
    \Delta v=\sqrt{\frac{GM_\star}{R_\mathrm{ring}}}\,\frac{\Delta p}{R_\mathrm{ring}}, 
\end{equation}
where \Mstar\ and $R_\mathrm{ring}$ are the stellar mass and ring radius, respectively. 
Given a distance of 445~pc \citep{2011Lombardi_2MASS_extinction_IV} and substituting the relation between $\Delta v$ and $\Delta p$ from Eq.~\ref{eq:dp_dv}, this equation can be approximated as:
\begin{equation}
    \label{eq:PV_CH3OH}
    M_\mathrm{*} = 3.56\,M_\odot \times \left ( \frac{R_\mathrm{ring}}{\mathrm{100~au}} \right )^3. 
\end{equation}
To further evaluate the size of the CH$_3$OH-rich feature, it is essential to constrain the stellar mass. 
This is addressed through YSO modeling, as discussed in Section~\ref{sec:model}.

\subsection{Molecular Diagnostics}
\label{sec:molec}

In addition to CO, C$^{18}$O, and CH$_3$OH, our ALMASOP observations detected several other molecular species, including CH$_3$CHO, HNCO, NH$_2$CHO, H$_2$CO, DCN, H$_2$S, and OCS. 
Table~\ref{tab:molec} provides details of the observed transitions, and Figure~\ref{fig:img_appx} presents the corresponding integrated intensity images.
For molecules with multiple detected transitions, as listed in Table~\ref{tab:molec}, we stacked the flux densities to improve the signal-to-noise ratio. 
As shown in Figure~\ref{fig:img_appx}, these molecules (CH$_3$OH, CH$_3$CHO, HNCO, NH$_2$CHO, H$_2$CO, DCN, H$_2$S, and OCS) exhibit a single-peaked morphology. 
Interestingly, although these molecules share this general spatial distribution, their kinematic properties display distinct patterns, suggesting different origins or underlying physical processes.

As shown in Figure~\ref{fig:PV_formulae_main}, the detected molecules can be broadly classified into two groups.
The first group, illustrated in Figure~\ref{fig:PV_formulae_main} (a)–(d), includes CH$_3$OH, CH$_3$CHO, HNCO, and NH$_2$CHO.
These molecules all display linear distributions in the PV diagrams haveing similar slopes.
This suggests that they are residing within the discovered CH$_3$OH-rich rotating feature, further indicating that this feature is not only abundant in CH$_3$OH but also in other COMs.

In contrast, as shown in Figure~\ref{fig:PV_formulae_main} (e)–(h), H$_2$CO, H$_2$S, OCS, and DCN exhibit a different kinematic signature from the COMs above.
First, at low velocities ($|\Delta v| < 2$ km s$^{-1}$), they all display extended emission ($|\Delta p| > 0\farcs{5}$), suggesting that these molecules are residing in the envelope. 
Meanwhile, in the inner region ($|\Delta p| < 0\farcs{5}$), the observed velocities significantly exceed the PV boundary derived from C$^{18}$O, and the emission shows two distinct peaks, one red-shifted and one blue-shifted. 
Such a pattern suggests that these molecules are located closer to the center.
If C$^{18}$O traces perfectly the envelope, these molecules also reside within the disk. 




\section{Young Stellar Object Modeling}
\label{sec:model}

In this section, we present the young stellar object (YSO) modeling.
A typical YSO model consists of a central protostar, a rotating disk, an infalling envelope, and an outflow cavity.
Developing a complete model requires numerous parameters, including the mass, temperature, and radius of the central protostar; the infall rate and size of the envelope; the mass, size, and shape of the disk; and the shape of the outflow cavity.
We begin by deriving some of these parameters directly from our observations.
For the remaining parameters, we perform spectral energy distribution (SED) fitting using a grid of models.


\subsection{Outflow Cavity}
\label{sec:model:cav}

\begin{figure}[htb!]
\centering
\includegraphics[width=.95\linewidth]{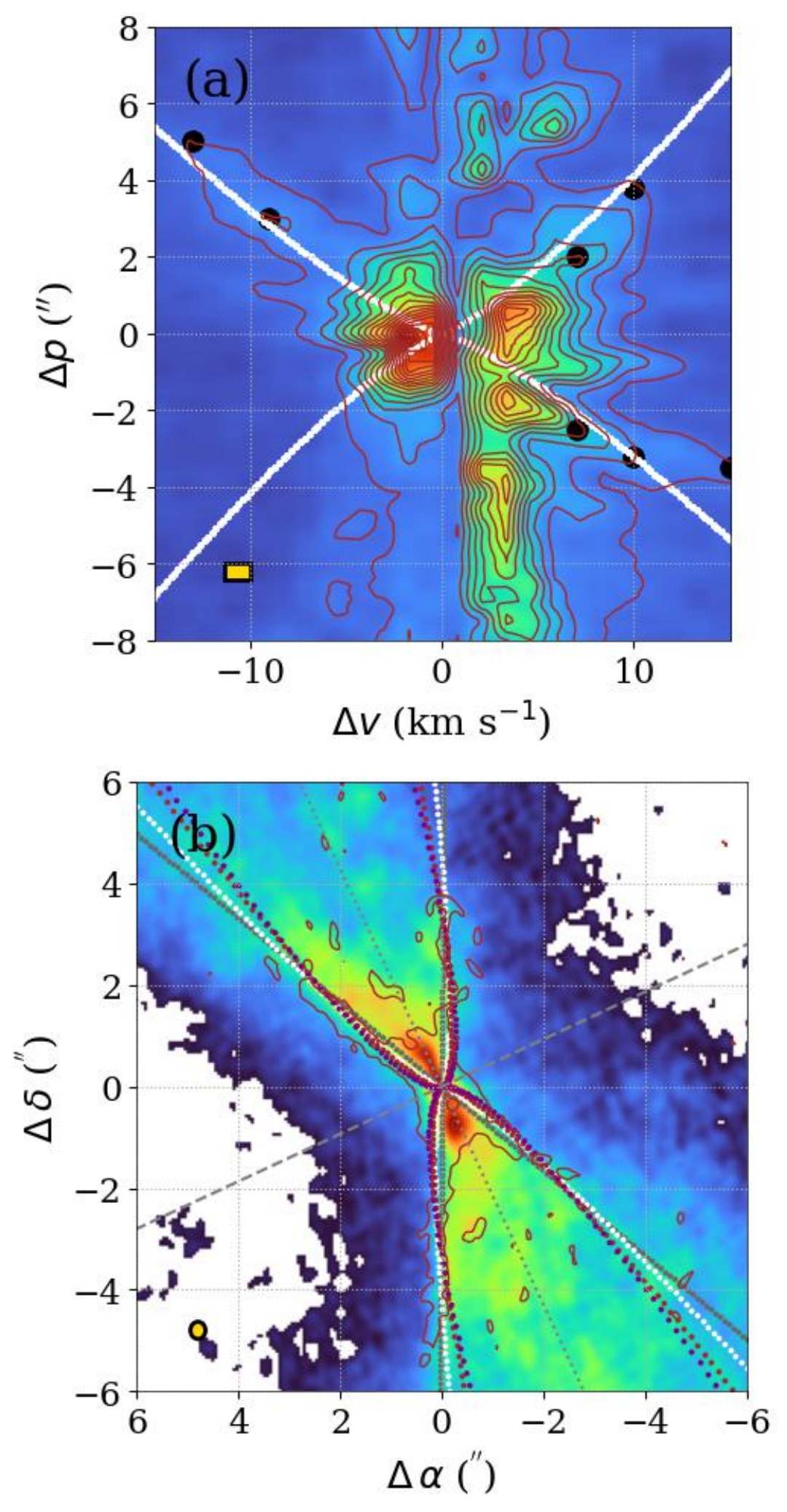}
\caption{\label{fig:best_cav} 
The PV diagram (a) and the integrated intensity map (b) of for CO $J=2-1$ transition. 
The PV diagram was extracted along the outflow axis, as illustrated by the dotted line in panel (b), and averaged by nine pixels (0\farcs{54}). 
The white curves illustrate our best-fit models. 
The black markers in panel (a) are the characteristic points we selected for fitting. 
The gold rectangles at the bottom left of the top panel represent the beam size and spectral resolution.
The contours in panel (b) illustrate the gradient of the integrated intensity map. 
The grey, white, brown, and purple dotted curves in panel (b) illustrate the best-fit cavity walls having wall exponent \bcav\ 1.00, 1.25, 1.50, and 1.75, respectively. 
In our modeling, we adopted the wall having \bcav$=1.25$ (white), as it is best-matching the cavity wall illustrated by the gradient map. 
}
\end{figure}

We start by evaluating the inclination angle ($\varphi$) and the parameters describing the outflow cavity walls using the CO $J=2-1$ observations.
The molecular CO outflow cavity was modeled as a radially expanding shell driven by an underlying wide-angle wind \citep{2000Lee_jet_wind_model}.
First, the velocity components along the $R$ and $z$ directions can be expressed using a velocity factor  ($\Tilde{v}$) as:
\begin{equation}
    v_\mathrm{R} =\Tilde{v} \, R
\end{equation}
and 
\begin{equation}
    v_\mathrm{z} =\Tilde{v} \, z. 
\end{equation} 
Considering an inclination angle ($\varphi$), the maximum projected blue- and red-shifted velocities along the line of sight at any point on the outflow axis $(R, z)$ is:
\begin{equation}
    v_\mathrm{proj}^\mathrm{blue} = v_\mathrm{R}\sin\varphi + v_\mathrm{z}\cos\varphi, 
\end{equation}
and 
\begin{equation}
    v_\mathrm{proj}^\mathrm{red} = -v_\mathrm{R}\sin\varphi + v_\mathrm{z}\cos\varphi
\end{equation}
and the projected height of the cavity wall is:
\begin{equation}
    z_\mathrm{proj} = z\sin\varphi. 
\end{equation}

To determine the the inclination angle ($\varphi$) as well as the velocity factor  ($\Tilde{v}$), we selected characteristic PV data points from the CO $J=2-1$ PV diagram along the outflow axis, as shown in Figure \ref{fig:best_cav} (a). 
The characteristic PV data points were selected to trace the arm-like pattern, which is a feature of the wide-angle wind model \citep{2000Lee_jet_wind_model}. 
The $z_\mathrm{proj}$ in the model is the $\Delta p$ in the PV diagram, and the $v_\mathrm{proj}^\mathrm{blue}$ and $v_\mathrm{proj}^\mathrm{red}$ are the negative and positive boundaries of $\Delta v$ at a given $\Delta p$. 
We fit the outflow model parameters by minimizing the $\chi^2$ value at each $\Delta v$ of the selected characteristic points in Figure~\ref{fig:best_cav}. 

Our best-fit parameters obtained through this process are $\varphi=93^{\circ}$ and $\Tilde{v}=1.7$ km au$^{-1}$s$^{-1}$. 
The angle of inclination of $93^{\circ}$ suggests that the system is nearly edge-on, with the northern lobe slightly tilted toward Earth (blue-shifted).
In the following, we ignore this tiny inclination and assume that the system is purely edge-on. 

Next, the cavity wall can be described in cylindrical coordinate $(R, \phi, z)$ with two parameters, $a_\mathrm{cav}$ and $b_\mathrm{cav}$, as:
\begin{equation}
    \label{eq:cav_wall}
    (\frac{z}{\mathrm{1~au}}) = a_\mathrm{cav}\,(\frac{R}{\mathrm{1~au}})^{b_\mathrm{cav}}. 
\end{equation}
These parameters of the wall were estimated using the gradient map of the CO integrated intensity (moment-0) image. 
We computed the sum of the gradient map values along paths defined by a set of cavity walls, each characterized by sampled parameters \bcav\ and \acav. 
The combination yielding the highest total (i.e., best-matching between the model and the observed gradient structure) was selected as the best-fit solution.
The parameter \bcav\ was sampled at three values: 1.00, 1.25, 1.50, and 1.75.
For each value of \bcav, \acav\ was chosen such that the radius $R$ spans from 1\arcs to 2\arcs at a height of $z=3\arcs$, assuming a distance of 445 pc, adopting Eq.~\ref{eq:cav_wall}.
The best-fit values, having the maximum sum, are \bcav$=1.25$ with \acav$=0.4$, as shown in Figure~\ref{fig:best_cav} (b). 
The derived exponent of 1.25 is smaller than the commonly assumed value of 1.5 \citep[e.g., ][]{2016Furlan_HOPS}, and the resulting wide cavity suggests that this source is more likely more evolved \citep[e.g., ][]{2023Hsieh_outflow,2024Dunham_outflow}.


\begin{figure}[htb!]
\centering
\includegraphics[width=.95\linewidth]{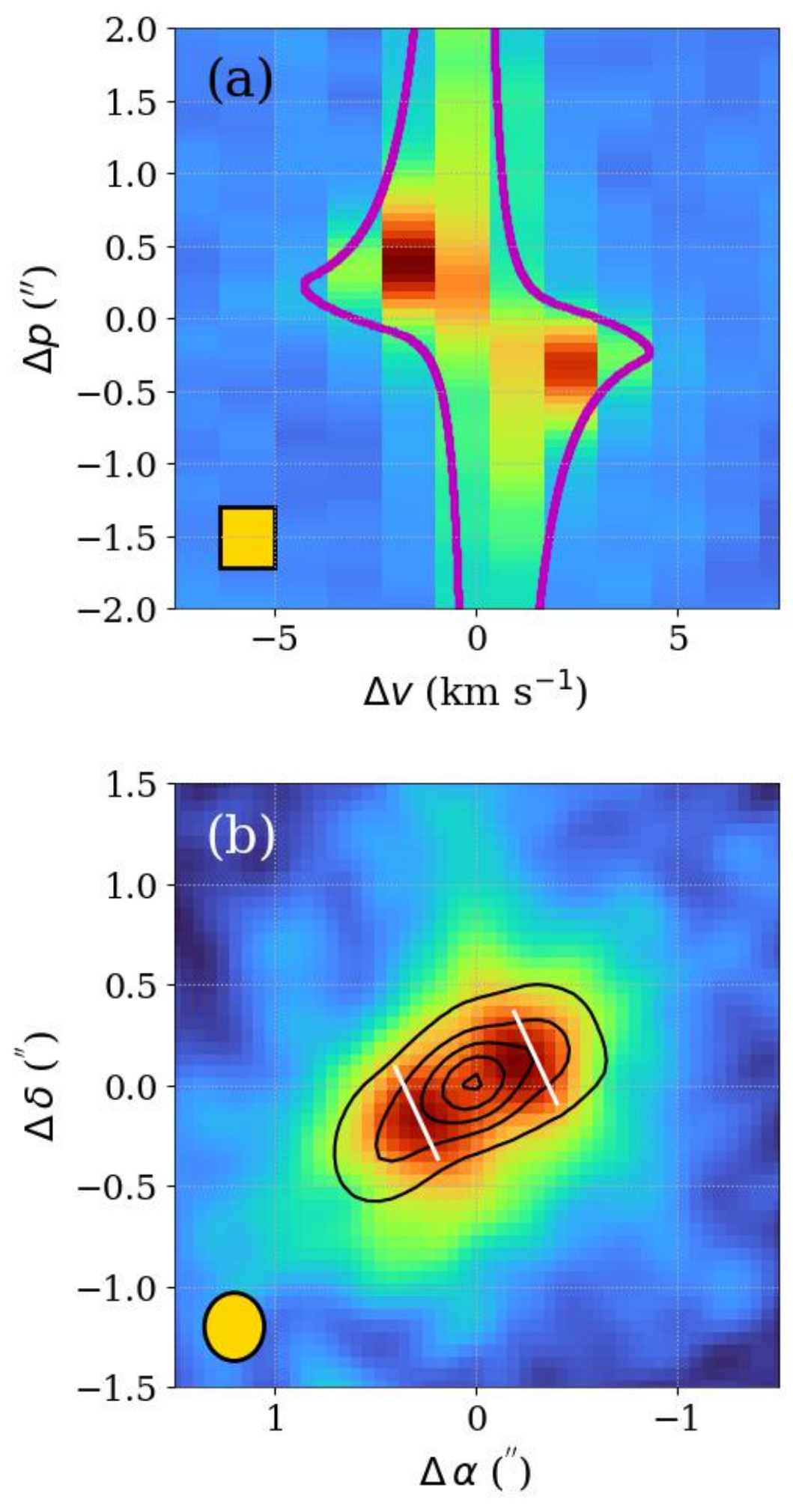}
\caption{\label{fig:best_env} 
The PV diagrams for C$^{18}$O (a) and the dust continuum overlaid on the C$^{18}$O integrated intensity (moment-0)} maps. 
The magenta curves in the top panel illustrate our best-fit models and the white lines in the bottom panel illustrate the disk size. 
The gold rectangle at the bottom left of the top panel represents the beam size and spectral resolution.
\end{figure}

\subsection{Envelope-disk System}
\label{sec:model:env-disk}

Two envelope models are commonly used to describe the dynamics of protostellar systems.
The first is the Ulrich-Cassen-Moosman (UCM or CMU) model, which characterizes a three-dimensional, azimuthally symmetric flow of rotating and infalling gas \citep[e.g.,][]{1976Ulrich_CMU,1981Cassen_CMU,2004Mendoza_UCM}.
The second is a simplified two-dimensional model introduced by \citet{2014Sakai_L1527_envelope}, based on the earlier work of \citet{1997Ohashi_IRAS043682557}, which assumes azimuthal symmetry and restricts all motion to the midplane \citep[e.g.,][]{1997Ohashi_IRAS043682557,2014Sakai_L1527_envelope,2014Oya_IRAS15398-3359,2022Oya_FERIA,2024Mori_model_UCM}.
In this study, we adopt the former. 
The radius at which the gravitational force balances the centrifugal force in the midplane is defined as the centrifugal radius ($R_\mathrm{C}$) and is expected to closely correspond to the disk radius.
Within the disk radius, the velocity is described by Keplerian rotation.

We derived the relations between the line-of-sight velocity and the position on the midplane for an edge-on envelope-disk system.
First, at the midplane of a UCM model under the spherical coordinate system ($r, \theta = 90^\circ, z=0$), the velocity components are:
\begin{align}
    v_r &= -\left (\frac{GM}{r} \right )^\frac{1}{2} \\
    v_\theta &= 0 \\
    v_\phi &= \left (\frac{GM}{r} \right )^\frac{1}{2}, 
\end{align}
where $G$ is the gravitational constant and $M$ is the central mass ($M_\mathrm{cen}=M_* + M_\mathrm{disk}$).
For an edge-on system, at a given point $(x, y)$, the velocity along the line of sight (x-direction, $v_\mathrm{los}$) is:
\begin{align}
    v_\mathrm{los} &= v_r \cos\phi - v_\phi \sin\phi \\
    &= -\sqrt{GM} \, \left ( x^2+y^2 \right )^{-\frac{3}{4}}\,(y+x), 
\end{align}
where $\phi$ is the polar angle (i.e., $\sin\phi=y/r$ and $\cos\phi=x/r$). 
Given a position along the midplane ($y$), the two local $v_\mathrm{los}$ extrema occur at:
\begin{equation}
x_\pm=\frac{-3\pm\sqrt{17}}{2}y\,\,\,.
\end{equation}
The two solutions represent the boundaries of the PV diagram at a given position y along the midplane. 
Each of them is replaced by $\pm\sqrt{R_C^2-y^2}$ if it is within the disk outer radius setting at the centrifugal radius $R_C$. 

We assumed that the C$^{18}$O emission is dominated by the extended envelope and derived the envelope parameters with its PV diagram. 
The PV diagram for C$^{18}$O were extracted along the major axis (i.e., perpendicular to the outflow axis). 
We measured the velocity boundaries (i.e., the maxima of the red and blue velocity shifts) at each position and fit the PV diagram by varying the centrifugal radius ($R_C$) and the central mass ($M_\mathrm{cen} = M_* + M_\mathrm{disk}$)
We then fit the envelope model by minimizing the $\chi^2$-value between the modeled and observed velocity boundaries. 

The best-fit parameters obtained are $R_C=145~\mathrm{au}$ and $M_\mathrm{cen}=1.50~M_\odot$. 
In this envelope-disk model, the centrifugal radius ($R_C$) is the boundary between the envelope and the disk. 
As shown in Figure~\ref{fig:best_env} (a), the modeled PV diagram (magenta curve) is well-matching the shape of the C$^{18}$O PV diagram. 
In Figure~\ref{fig:best_env} (b), we show the modeled disk size (white lines) and the observed 1.3~mm dust continuum (black contours) on the observed C$^{18}$O integrated intensity map (rasters). 

\begin{deluxetable*}{lllll}[tbp!]
\label{tab:params}
\caption{Parameters for YSO Model Grid}
\tablehead{
\colhead{Symbol} & \colhead{Description} & \colhead{Best-fit value} & \colhead{Sampled Value}
}
\startdata
\multicolumn{4}{c}{Protostar}\\
$M_*$ & Stellar mass & 1.3 \Msun & $1.50M_\odot-M_\mathrm{disk}$\\
$T_*$ & Stellar effective temperature & 7000 K & 5000, 6000, 7000, 8000~K\\
$r_*$ & Stellar radius & 2 \Rsun & 1, 2, 3, 4, 5, 6~$R_\odot$\\
\hline
\multicolumn{4}{c}{Envelope}\\
$r_\mathrm{env}$ & Envelope outer radius & 10,000~au \\
$r_\mathrm{min}$ & Inner radius & $R_\mathrm{sub}$ \\
$\dot{M}_\mathrm{env}$ & Mass infall rate & $10^{-4.0}$ \Msunyr & $10^{-5.0}$, $10^{-4.5}$, $10^{-4.0}$, $10^{-3.5}$ $M_\odot$~yr$^{-1}$ \\
\hline
\multicolumn{4}{c}{Disk}\\
$M_\mathrm{disk}$ & Disk mass & 0.20 \Msun & 0.05, 0.10, 0.15, 0.20, 0.25 \Msun\\
$R_\mathrm{disk}$ & Disk outer radius & 145~au \\
$R_\mathrm{min}$ & Inner radius & $R_\mathrm{sub}$ \\
$\dot{M}_\mathrm{disk}$ & Mass accretion rate & $10^{-6.0}$ \Msunyr & $10^{-7.0}$, $10^{-6.5}$, $10^{-6.0}$, $10^{-5.5}$ $M_\odot$~yr$^{-1}$\\
$A_\mathrm{disk}$ & Radial exponent in disk density law & 2.25 \\
$B_\mathrm{disk}$ & Vertical exponent in disk density law & 1.25 \\
$R_\mathrm{trunc}$ & Magnetospheric truncation radius & $3R_*$ \\
$f_\mathrm{spot}$ & Fractional area of the hot spots & 0.01 \\
\hline
\multicolumn{4}{c}{Cavity}\\
$b_\mathrm{cav}$ & Cavity wall exponent & 1.25 &  \\
$a_\mathrm{cav}$ & Cavity opening angle factor & 0.4 & 
 \\
\enddata
\tablecomments{
The central mass ($M_\mathrm{cen} = M_* + M_\mathrm{disk}$) and centrifugal radius (\Rc=\RdiskMAX) are derived from fitting the C$^{18}$O PV diagram. 
The cavity wall component (\bcav) and opening angle factor (\acav) are derived from fitting the CO PV diagram. 
The dust grain models are ``www003'' \citep{2002Wood_dust} for the dense region of the disk, ``kmh'' \citep{1994Kim_dust} for the diffuse region of the disk and the outflow, and ``r400\_ice095'' \citep{2003Whitney_HOCHUNK_II} for the envelope, referring to \citet{2013_Whitney_HOCHUNK}.  
}
\end{deluxetable*}


\subsection{Spectral Energy Distribution (SED) Fitting}
\label{sec:model:SED}

For the remaining parameters, we applied spectral energy distribution (SED) fitting on the source, with the photometric datapoints collected from \citet{2020Dutta_ALMASOP} and shown in Table \ref{tab:SED}. 
The surveys include WISE \citep{2010Wright_SED_WISE}, AKARI \citep{2010Ishihara_SED_AKARI_IRC}, and JCMT SCUBA \citep{2008Francesco_JCMTS_submm}. 
The covered wavelength ranges from 3.4 \micron\ to 850 \micron. 

\begin{figure}[htb!]
\centering
\includegraphics[width=.95\linewidth]{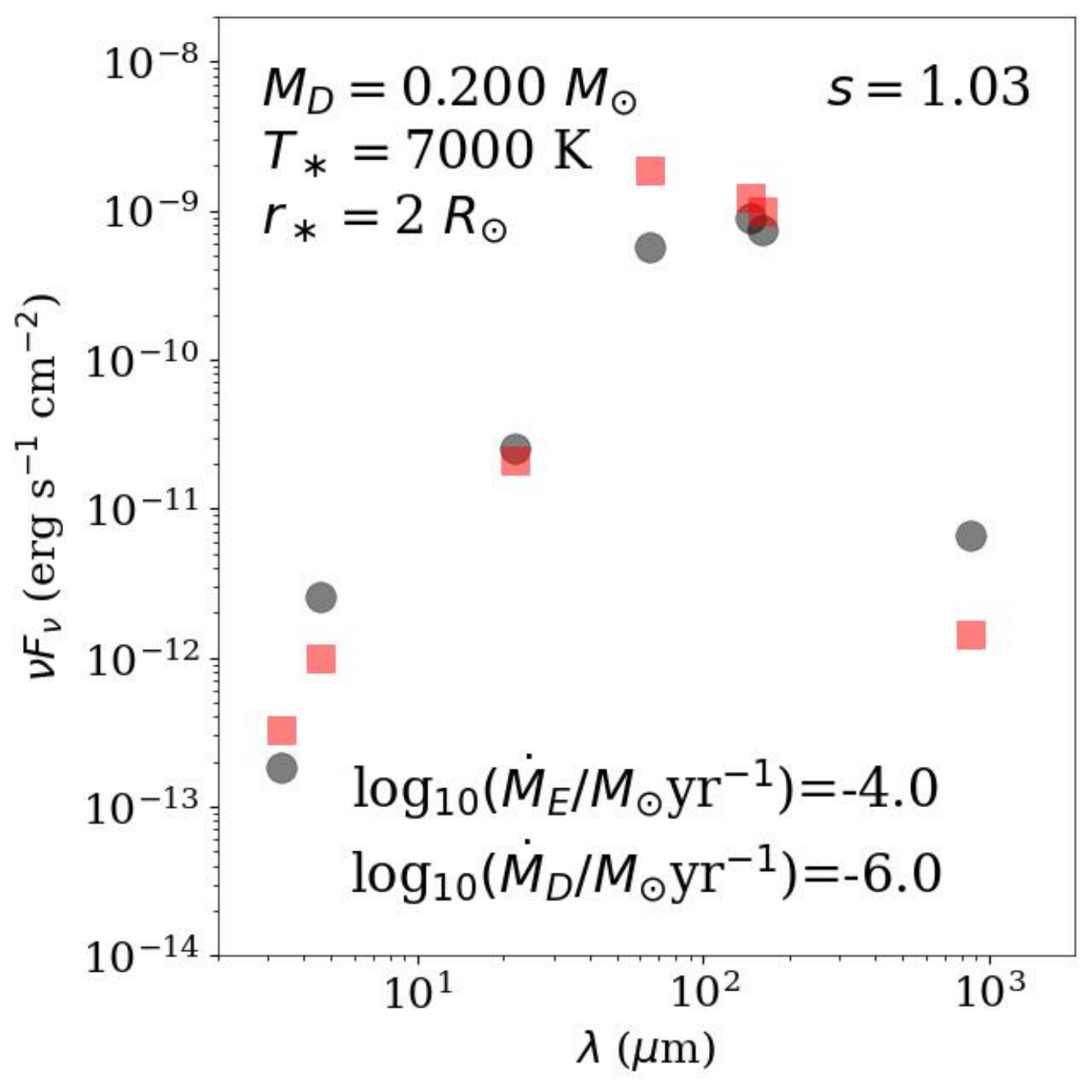}
\caption{\label{fig:SED_best} 
The observed and best-modeled SED. 
Black markers represent the observed data points, while red markers indicate the modeled data points. 
All sampled parameters and the corresponding $\chi^2$ values are labeled in the panel. 
}
\end{figure}

The tool used for calculating the SEDs of the YSO model grid is \texttt{HOCHUNK} \citep{2003Whitney_HOCHUNK_I, 2003Whitney_HOCHUNK_II}.
Descriptions of the parameters are provided in Table \ref{tab:params}.
The model grid was constructed by sampling five key parameters: stellar temperature (\Tstar), stellar radius (\rstar), envelope mass infall rate (\MDOTenv), disk mass (\Mdisk), disk mass accretion rate (\MDOTdisk).
The stellar mass (\Mstar) was also sampled, as the central mass was fixed at $M_\mathrm{cen} = M_* + M_\mathrm{disk}$.
To reduce computation time, certain parameters were manually assigned.
The inner radii of both the disk and envelope were set to the dust sublimation radius (\Rsub).
The disk height was limited to the disk scale height, which was defined as the hydrostatic equilibrium at the disk's inner radius.
The dust grain models used were ``www003'' \citep{2002Wood_dust} for the dense regions of the disk, ``kmh'' \citep{1994Kim_dust} for the outflow and the diffuse regions of the disk, and ``r400\_ice095'' \citep{2003Whitney_HOCHUNK_II} for the envelope, following \citet{2013_Whitney_HOCHUNK}.

\texttt{HOCHUNK} generates the density profile, computes the temperature profile, and produces the SEDs for the YSO models.
For each SED, we additionally incorporated a scaling factor ($s$) ranging from 0.90 to 1.10 from \citet{2016Furlan_HOPS}. 
To evaluate the goodness of fit, we adopted the method from \citet{2007Robitaille_sedfitter}, calculating the $\chi^2$ value between the observed and modeled photometric data points in logarithmic space.
For each model, we first determined the best-fit values of $s$.
We then ranked the combinations of YSO models and scaling factor ($s$) based on their $\chi^2$ values, identifying the best matches to the observed data. 

The combination of sampled parameters having minimum $\chi^2$ value is: \Tstar$=7000$~K, \rstar$=2$\Rsun, \MDOTenv$=10^{-4.0}$\Msunyr, \Mdisk$=0.2$~\Msun, and \MDOTdisk$=10^{-6.0}$~\Msunyr. 
Figure~\ref{fig:SED_best} shows the corresponding SED overlaid with the observations. 
The derived stellar mass is $M_\mathrm{cen} - M_\mathrm{disk} = 1.30 M_\odot$, suggesting that this core is close to an intermediate-mass protostellar core.
Substituting the stellar mass (1.30 \Msun) in Eq. \ref{eq:PV_CH3OH}, the estimated COM-rich feature size is around 71~au, within the disk size. 
Despite the disk mass (0.20 \Msun) is not much smaller than the stellar mass (1.30 \Msun), the uncertainty to the estimated size is less than 5~au. 


\begin{figure*}[htb!]
\centering
\includegraphics[width=.99\linewidth]{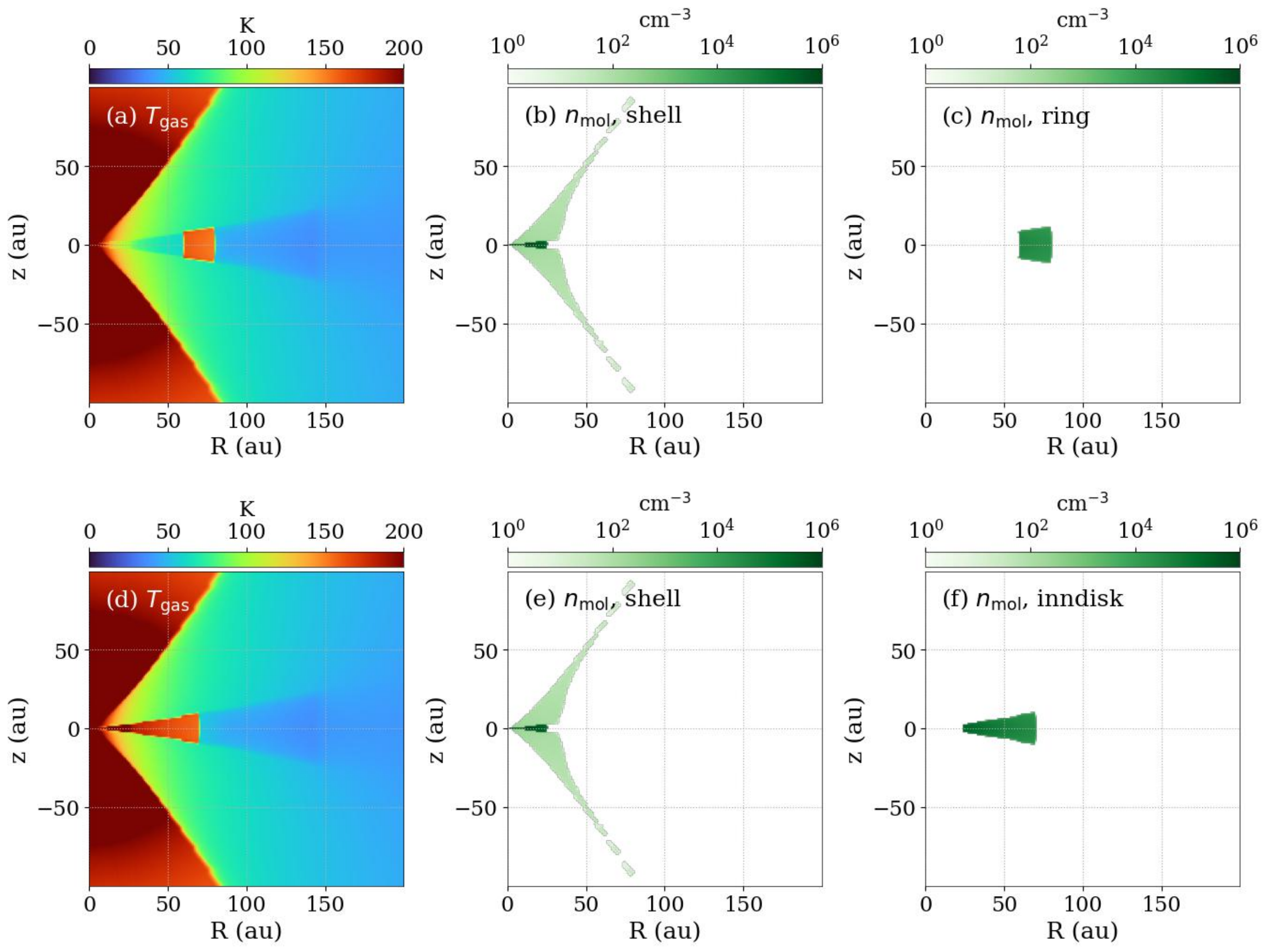}
\caption{\label{fig:phys_sparx_best} 
The temperature profiles, panel (a) and (c), and the resulting CH$_3$OH extent models used for image synthesis.
The shell structures shown in panels (b) and (e), located near the outflow cavity and encompassing both the inner envelope and the innermost disk, correspond to the warm region predicted by the YSO model simulation. 
In addition, the square in panel (c) and the wedge in panel (f), both positioned around 70 au, were introduced based on the YSO modeling results to represent a ring and an inner disk in 3D space. 
Note that this inner disk does not include the innermost warm disk region originally produced by the YSO model. 
}
\end{figure*}


\subsection{Image Synthesis}
\label{sec:xclass}

\begin{figure*}[htb!]
\centering
\includegraphics[width=.99\linewidth]{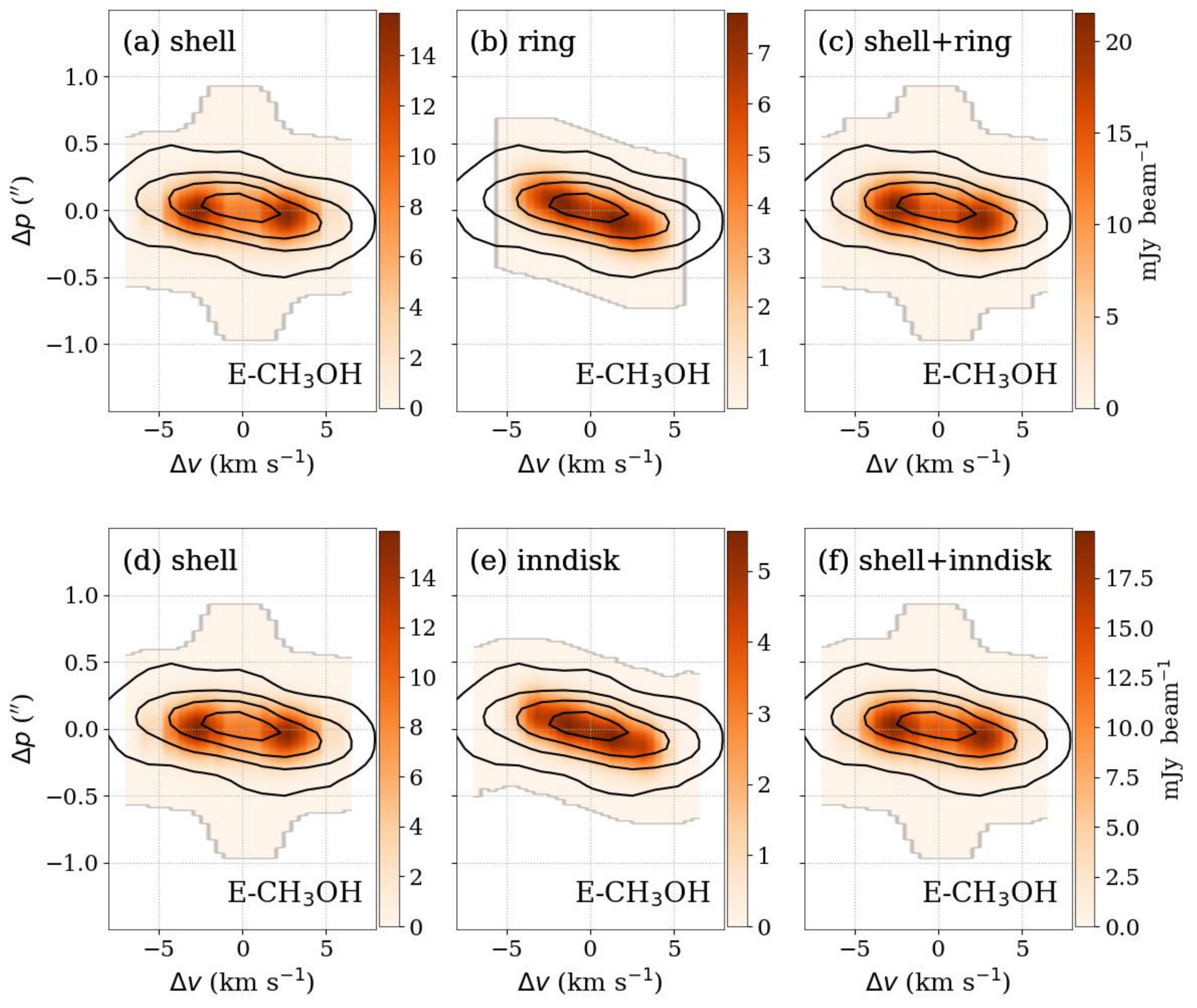}
\caption{\label{fig:PV_sparx_main} 
The modeled PV diagrams (rasters) are overlaid with the observed data (black contours).
The top and bottom rows correspond to the temperature profiles with additionally introduced warm ring and warm inner disk, respectively. 
In the top panels, from left to right, CH$_3$OH is assumed to reside in the shell alone (``shell"), the warm ring (``ring"), and both of them (``shell+ring").
In the bottom panels, from left to right, CH$_3$OH is assumed to reside in the shell (``shell"), the warm inner disk (``inndisk''), and both warm regions combined (``inndisk+shell").
For a clearer illustration of these regions, see Figure~\ref{fig:phys_sparx_best}.
}
\end{figure*}

Next, we aimed to discover the CH$_3$OH extent model capable of producing the rotating feature in the observeations.
We adopted the temperature maps exported by \texttt{HOCHUNK} of the best-fit YSO model and used \texttt{SPARX}\footnote{https://charms.asiaa.sinica.edu.tw/sparx/} (Simulation Package for Astronomical Radiative Xfer) to produce synthetic image cubes with the density, velocity, and temperature, assuming local-thermodynamic-equilibrium (LTE) and a relative abundance $X$ of $10^{-6}$ \citep{2015Boogert_ice_review}.
The targeted transition, CH$_3$OH $5_{1,4}$ -- $4_{2,4}$, has an upper energy level of 45~K.

Following the assumption that the gas-phase CH$_3$OH molecules are thermally desorbed, a temperature of 100~K was required to reach this abundance.
However, while gas-phase CH$_3$OH is typically expected to arise from thermal desorption, the imported physical structure did not include a sufficiently warm feature at radius $\sim70$~au due to the disk shadowing effect \citep[see also, e.g.,][]{2022Nazari_model_diskshadow,2023Hsu_ALMASOP}. .
To address this, we introduced a localized heating effect within the disk at radii between 60 and 80 au, raising the temperature for 100 K. 
As shown in Figure \ref{fig:phys_sparx_best} (a), two regions exceed the ice sublimation temperature of 100 K, where we assume the presence of CH$_3$OH.
One region, shown in Figure~\ref{fig:phys_sparx_best} (b), is an inner shell near the cavity wall, encompassing both the inner part of the envelope and the innermost part of the disk, and is primarily heated by the central protostar (``shell''). 
The other one, shown in Figure~\ref{fig:phys_sparx_best} (c), corresponds to the localized heating we introduced around 70 au in the midplane (``ring''). 
We tested three spatial distribution models, where CH$_3$OH was assumed to reside either exclusively in the warm ring (``ring''), in the warm shell (``shell''), or in both regions (``ring+shell''), as marked in Figure~\ref{fig:phys_sparx_best}.


\begin{figure*}[htb!]
\centering
\includegraphics[width=.95\linewidth]{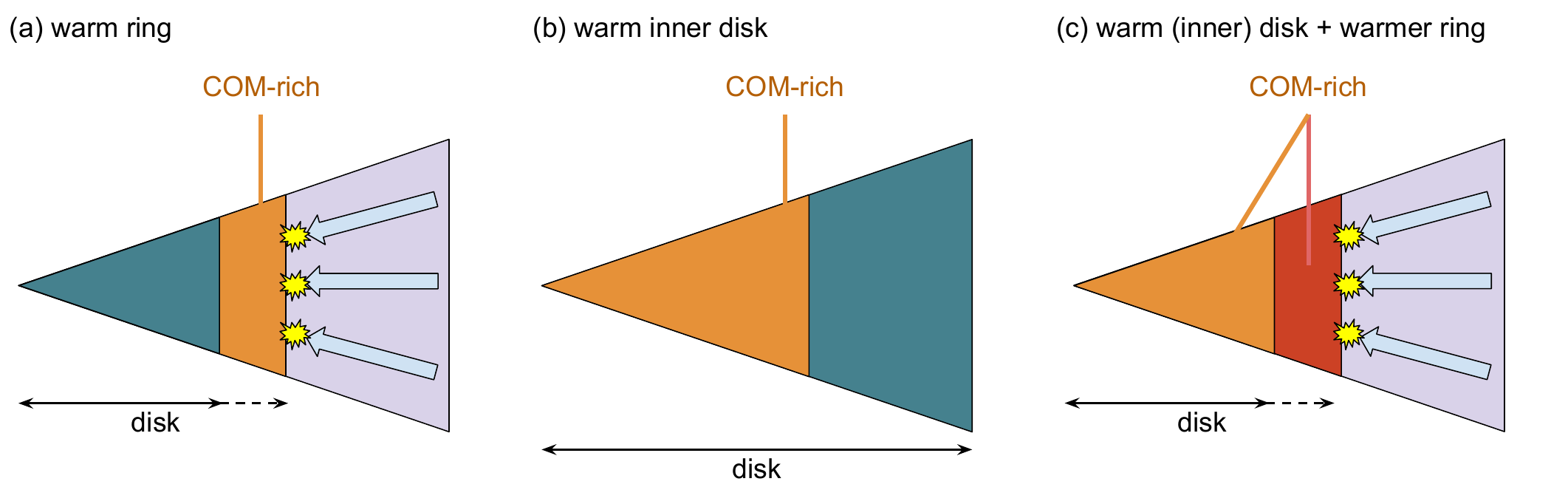}
\caption{\label{fig:schematic} 
The schematic illustrates potential scenarios for COM distribution.
The warm ring (a) can be explained by shocks from accreting material. 
The warm inner disk (b) can be due to stellar outburst, stellar radiation, and disk-wind interaction. 
These scenarios are not mutually exclusive (c). 
}
\end{figure*}

Figure~\ref{fig:PV_sparx_main} (a)–(c) presents the simulated PV diagrams (orange rasters) for these three extent models overlaid with the observed PV diagrams (black contours).
In the ``shell'' model, CH$_3$OH molecules reside in the inner warm region, producing velocity profiles with two peaks at $\Delta p\sim0\arcs$, corresponding to the blue- and red-shifted infall motions. 
In contrast, as expected, the ``ring'' model exhibits a clear velocity gradient, closely matching our observations. 

Despite these efforts, some discrepancies remain between the modeling results and observations.
First, the ``ring'' model still tentatively exhibits two peaks in the PV diagram, whereas the observations show only a single peak. 
Second, if CH$_3$OH is ubiquitous within the icy mantles of dust grains, a model where CH$_3$OH exists in all warm regions--the ``ring+shell'' model--should better match the observations.
However, the ``shell'' model produces stronger CH$_3$OH emission than the ``ring'' model, making the ``ring+shell'' model more similar to the ``shell'' model rather than aligning with the observed data.
A more refined disk model and chemical simulation will be necessary to further address these discrepancies.

As an unresolved disk could also possibly exhibit a linear pattern in PV diagram, we further introduce another heating effect, which rises the temperature for 100~K within the disk inside a radius of 70 au. 
The resulting temperature profile and the two warm regions are shown in~\ref{fig:phys_sparx_best} (d)-(f). 
The resulting PV diagrams are shown in Figure~\ref{fig:PV_sparx_main} (d)-(f). 
Overall, the ``inndisk'' model produces results very similar to those of the ``ring'' model, although the two-peak signature in the latter is slightly more discernible.
hese results suggest that the observed rotating COM-rich feature could potentially be an unresolved Keplerian disk.


\section{Discussion}
\label{sec:Disc}

\subsection{The Origin of the Observed Rotating COM-rich Featrue}


As shown in \S~\ref{sec:xclass}, the observed COM-rich rotating feature may be either a warm ring (``ring'') or a compact warm disk (``inndisk''). 
In this section, we discuss possible scenarios for the origins of the observed COM-rich feature for the two distinct cases. 

The presence of a warm ring could be explained by the ``accretion shock'' scenario, as shown in Fig.~\ref{fig:schematic} (a). 
The ``accretion shock" occurs at the boundary between the envelope and the outer disk, where material arrives from the former to the latter. 
Observational support for accretion shocks was first provided by \citet{2002Velusamy_L1157}, based on CH$_3$OH emission detected at the infall-disk interface in L1157.
In L1527, \citet{2014Sakai_L1527_envelope} identified a sharp change in chemistry at the disk edge, possibly due to the localized heating caused by accretion shocks. 
In IRAS 16293-2422 A, the accretion shock mechanism was  proposed to explain the narrowly distributed COM emission \citep{2016Oya_IRAS162932422} and the steep rise of the rotational temperature of H$_2$CS \citep{2020Oya_IRAS16293-2422-A_few_au}. 
Also, a ring-like warm structure in B335 was also suggested to be caused by the accretion shocks  \citep{2022Okoda_B335_few_au}. 
In our case, the radius of the COM-rich feature ($\sim$70 au) is not comparable to the modeled disk size ($\sim$140 au), but rather about half of it. 
This suggests that the modeled disk size may be overestimated if the COM-rich feature is caused by accretion shocks.
The disk size was derived by fitting the C$^{18}$O PV diagram under the assumption that C$^{18}$O perfectly traces the envelope. 
The observed discrepancy may indicate that C$^{18}$O emission is weaker in the inner envelope, thereby failing to reveal the high-velocity component in the inner region clearly.
It is also possible that the disk edge does not lie at the centrifugal radius, but rather at half of it.
This behavior is described by the ``SB model,'' in which the disk edge is located at the so-called centrifugal barrier, defined as half the centrifugal radius \citep[e.g.,][]{2014Sakai_L1527_envelope,2014Oya_IRAS15398-3359,2022Oya_FERIA}.
In our case, the modeled centrifugal radius is 145 au, placing the centrifugal barrier at approximately 70 au, comparable to the size of the observed COM, rich feature.
Alternatively, we note that, in contrast, if the COM-rich feature is indeed a ring inside the disk, this could possibly a localized desorption mechanism inside the disk. 


In contrast, the presence of a warm inner disk could be explained by a stellar outburst, which has been proposed to account for the presence of COMs in the V883 Ori disk \citep{2018vantHoff_V883Ori_CH3OH, 2019Lee_V883-Ori_outburst, 2025Jeong_V883Ori_COM}. 
These outbursts result from episodic mass accretion events \citep[e.g.,][]{2014Audard_PPVI, 2023Fischer_review}, which can rapidly elevate the temperature of the inner disk and trigger the desorption of COMs. 
After such an event, the desorbed COM molecules can re-freeze onto icy grains when the temperature falls below the sublimation threshold. 
This re-freezing process can occur on timescales as short as one year \citep{2004Lee_chem_evolution, 2019Lee_V883-Ori_outburst}.
Alternatively, in the case of HH~212 where such stellar outburst was not observed, radiation from the central protostar, or interactions with disk winds have been proposed for the presence of warm disks \citep{2022Lee_HH212_stratification}. 

The accretion shock and the mechanism responsible for the warm disk may not be mutually exclusive, as illustrated in Figure~\ref{fig:schematic}(c). 
For instance, in HH~212, some COMs near the disk edge may be released by heat generated from accretion shocks \citep{2017Lee_HH212, 2017Tabone_HH212_cavity}, even though the presence of a warm disk has also been suggested \citep{2022Lee_HH212_stratification}. 
In such cases, spatially resolved measurements of temperature and/or kinematics are necessary to further constrain the underlying physical processes.



\subsection{Comparisons with the COM-rich Disks and Implications}

So far, only a handful of disks in protostellar cores have been reported to harbor COMs, incuding HH~212, V883 Ori, TW Hya, and DG Tau \citep{2016Codella_HH212_H2O_COM,2017Lee_HH212,2019Lee_HH212_COM_atm,2022Lee_HH212_stratification,2018vantHoff_V883Ori_CH3OH,2019Lee_V883-Ori_outburst,2025Jeong_V883Ori_COM,2025Fadul_V883Ori,2016Walsh_TWHya_CH3OH,2015Loomis_DMTau}. 
Among them, G192.12-11.10 shares multiple similarities with HH~212. 
Both sources are solar-like protostellar cores located in the Orion molecular cloud and host nearly edge-on disks.

A key distinction between G192.12-11.10 and HH~212 is their evolutionary stage.
HH~212 has been classified as a Class~0 protostellar core and is associated with a collimated molecular jet.
In contrast, our non-detection of a SiO jet detection suggests that it G192.12-11.10 more evolved, probably Class I \citep[e.g.,][]{2016Bally_review_jet}. 
The wide cavity wall of the observed CO outflow also support that G192.12-11.10 is more evolved. 
The relatively low bolometric temperature  (\Tbol$=44\pm15$ K) calculated by \citet{2020Dutta_ALMASOP} may be explained by the nearly edge-on geometry. 
If the detected COMs exhibiting rotating motion in G192.12-11.10 indeed reside in the disk, as seen in HH~212, then this system represents a slightly later stage of disk chemistry at early phases (Class~0 and I).
Comparing these two protostellar cores will provide critical insights into the evolution of COMs within protostellar disks over time.


\subsection{Molecular Diagnostics and Implications}

\begin{figure}[htb!]
\centering
\includegraphics[width=.95\linewidth]{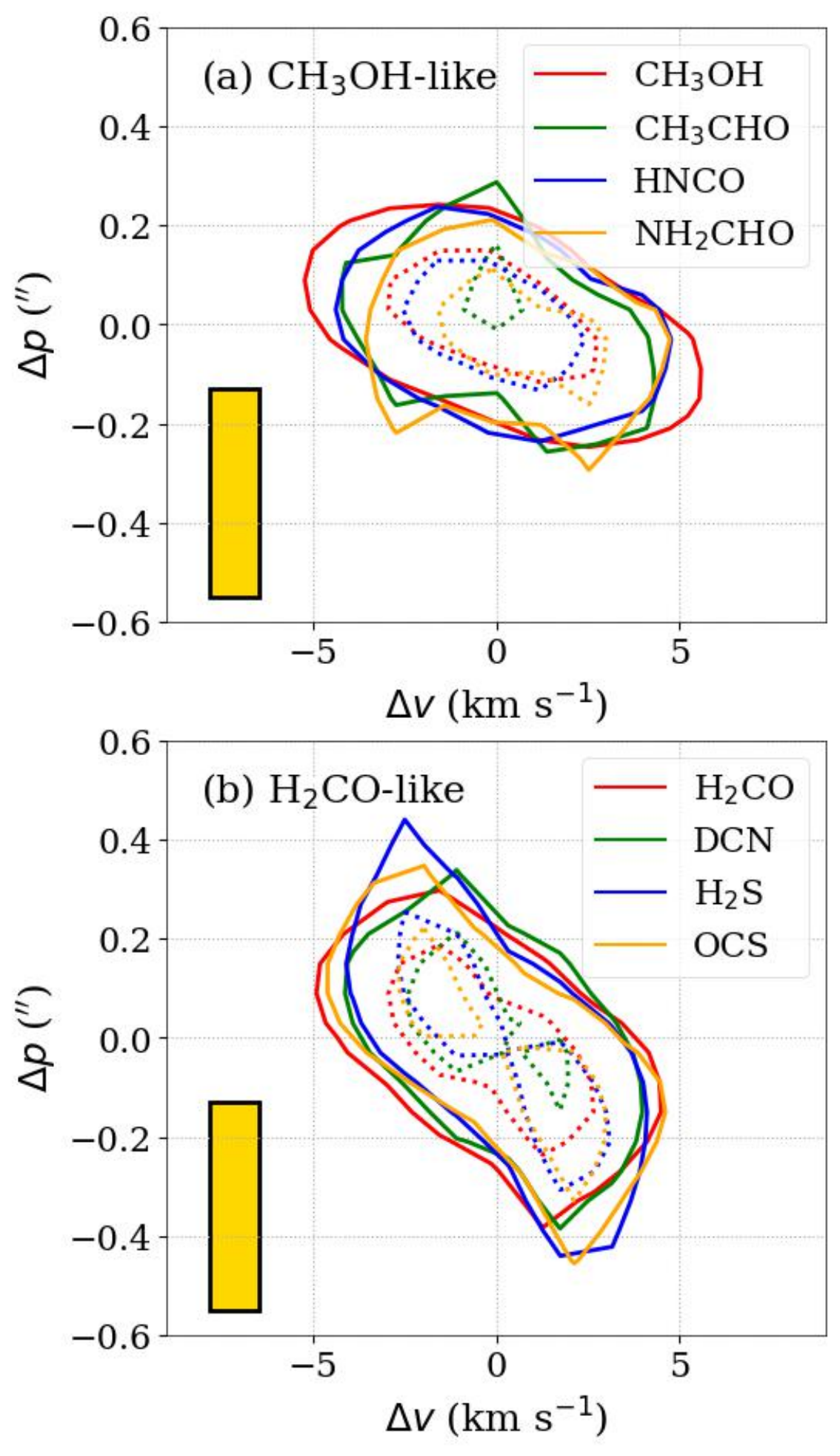}
\caption{\label{fig:PV_stratification} 
The PV diagrams of the selected molecular species. 
The solid and dotted contour levels are 50\% and 85\% of the peak flux density, respectively. 
The gold rectangles at the bottom left of each panel represent the beam size and spectral resolution.
}
\end{figure}

The stratified distribution of molecules was previously observed in the protostellar disk of HH~212 and the protoplanetary disk of DG Tau \citep{2022Lee_HH212_stratification, 2020Podio_ALMADOT_interplay}.
Accurately measuring the spatial extent of different molecules not only constrains the chemical network but also provides insights into the physical properties of the disk.
For instance, the agreement between molecular radii and their binding energies in HH~212 suggests that the detected species were thermally desorbed from ice.

We attempt to investigate the variations between the detected molecules in G192.12-11.10. 
As presented in \S \ref{sec:molec} and shown in Figure \ref{fig:PV_formulae_main}, the detected molecules can be broadly categorized into two groups: CH$_3$OH-like and H$_2$CO-like. 
In Figure~\ref{fig:PV_stratification}(a) and (b), we overlay the PV diagrams of the molecules for the two groups individually.
As seen in Figure~\ref{fig:PV_stratification}(a), no significant slope differences are identified among the CH$_3$OH-like molecules, implying that they reside in rotating features of comparable radii.
However, the limited spatial and spectral resolution prevents a more detailed examination of potential variations.

In contrast, as displayed in Figure~\ref{fig:PV_stratification}(b), the differences between the H$_2$CO-like molecules (H$_2$CO, DCN, H$_2$S, and OCS) can be tentatively identified. 
H$_2$CO and OCS appear to be have larger $|\Delta v|$ at $|\Delta p| < 0\farcs{2}$, compared with DCN and H$_2$S. 
Considering that H$_2$CO reaches comparable velocities to CH$_3$OH at $\lvert \Delta p \rvert < 0\farcs{25}$, as shown in Figure~\ref{fig:PV_formulae_main}, it is possible that both H$_2$CO and OCS also reside within the COM-rich feature in addition to the envelope.

Among these molecules, H$_2$S serves as a precursor to OCS.
Laboratory experiments suggest that ice-phase OCS can efficiently form from H$_2$S-ice mixtures through photodissociation by UV photons \citep{2008Ferrante_model_OCS,2010Garozzo_model_sulfur,2015Chen_H2S_OCS}.
Consistently, our observations show that OCS resides in the inner regions compared to H$_2$S, where it is exposed to more intense stellar irradiation from the central protostar.
A similar trend is observed in L1527, another protostellar core, where OCS is found to be more compact and concentrated near the protostar relative to other sulfur-bearing molecules such as SO, CS, and H$_2$CS \citep{2024Zhang_L1527_sulfur}.
We note that OCS was also observed in various environment, such as outflow front and circumbinary disk \citep{2007Benedettini_L1157,2020Takakuwa_L1551}. 

Finally, our study identifies both the kinematic difference between NH$_2$CHO ad H$_2$CO and the kinematic similarity between NH$_2$CHO and HNCO. 
Similarly, based on the different morphology and kinematics of NH$_2$CHO and H$_2$CO, \citet{2022Lee_HH212_stratification} questioned whether H$_2$CO could serve as a precursor to NH$_2$CHO.
Additionally, the consistent spatial extent, kinematics, and temperature of HNCO and NH$_2$CHO led them to suggest that HNCO is a daughter species of desorbed NH$_2$CHO.
Our findings support the conclusions of \citet{2022Lee_HH212_stratification}.


\section{Conclusions}
\label{sec:Conclusions}

Under the ALMASOP project, we investigated the morphology and kinematics of CO isotopologues and COMs in the protostellar core G192.12-11.10. Our key findings include:

\begin{enumerate}
    \item CO, C$^{18}$O, and CH$_3$OH appear to exhibit rotational motion and trace the outflow, envelope, and a rotating feature, respectively.  
    The CH$_3$OH-rich feature is also enriched in CH$_3$CHO, HNCO, and NH$_2$CHO
    Our observations do not reveal any significant differences in their kinematics and radial distribution.
    In contrast, H$_2$CO, H$_2$S, OCS, and DCN appear to have different nature, while H$_2$CO and OCS are possibly residing in the COM-rich rotating feature as well. . 

    \item We constructed a YSO model for the source by (1) fitting the CO and C$^{18}$O observations and (2) performing spectral energy distribution (SED) fitting.  
    The resulting image synthesis suggests that an additional heating mechanism may be necessary to reproduce the observed COM emission.  
    The modeling results suggest that the COM-rich feature has a size less than or comparable to the disk size. 

    \item We explore two possible scenarios for the COM-rich feature: a warm ring and a warm inner disk, noting that these scenarios are not mutually exclusive. A more refined envelope-disk model together with observations at higher spatial resolution and with more accurate temperature measurements, will be crucial for further constraining the physical structure.

    \item The protostellar core G192.12-11.10 exhibits several similarities to HH~212. 
    A thorough comparison of these two sources could offer valuable insights into the evolution of COMs in protostellar disks over time.

\end{enumerate}



\acknowledgments
This paper makes use of the following ALMA data: ADS/JAO.ALMA\#2018.1.00302.S. ALMA is a partnership of ESO (representing its member states), NSF (USA) and NINS (Japan), together with NRC (Canada), NSTC and ASIAA (Taiwan), and KASI (Republic of Korea), in cooperation with the Republic of Chile. The Joint ALMA Observatory is operated by ESO, AUI/NRAO and NAOJ.
This work made use of Astropy:\footnote{http://www.astropy.org} a community-developed core Python package and an ecosystem of tools and resources for astronomy \citep{astropy:2013, astropy:2018, astropy:2022}. 
S.-Y. H. and C.-F.L. acknowledge grant from the National Science and Technology Council of Taiwan (112-2112-M-001- 039-MY3).  
D.J.\ is supported by NRC Canada and by an NSERC Discovery Grant.
X.L. acknowledges the support of the Strategic Priority Research Program of the Chinese Academy of Sciences under Grant No. XDB0800303. 
The work of MGR is supported by the international Gemini Observatory, a program of NSF NOIRLab, which is managed by the Association of Universities for Research in Astronomy (AURA) under a cooperative agreement with the U.S. National Science Foundation, on behalf of the Gemini partnership of Argentina, Brazil, Canada, Chile, the Republic of Korea, and the United States of America.
L. B. gratefully acknowledges support by the ANID BASAL project FB210003
PS was partially supported by a Grant-in-Aid for Scientific Research (KAKENHI Number JP22H01271 and JP23H01221) of JSPS. 
W.K. was supported by the National Research Foundation of Korea (NRF) grant funded by the Korea government (MSIT) (RS-2024-00342488).
MJ acknowledges the support of the Research Council of Finland Grant No.
348342.
D.S. acknowledges the support from  Ramanujan Fellowship (SERB)  and PRL, India. 

\software{
Astropy \citep{astropy:2013, astropy:2018, astropy:2022},
\texttt{CASA} \citep{casa:2022},
\texttt{CARTA}  \citep{2021Comrie_CARTA}, 
\texttt{HOCHUNk} \citep{2003Whitney_HOCHUNK_I,2003Whitney_HOCHUNK_II}, 
\texttt{SPARX} https://charms.asiaa.sinica.edu.tw/sparx/), 
\texttt{XCLASS} \citep{2017Moller_XCLASS}, 
}

\clearpage
\appendix

\restartappendixnumbering

\section{Molecular Diagnostics \label{appx:molec}}
\resetapptablenumbers

\begin{figure*}[htb!]
\centering
\includegraphics[width=.95\linewidth]{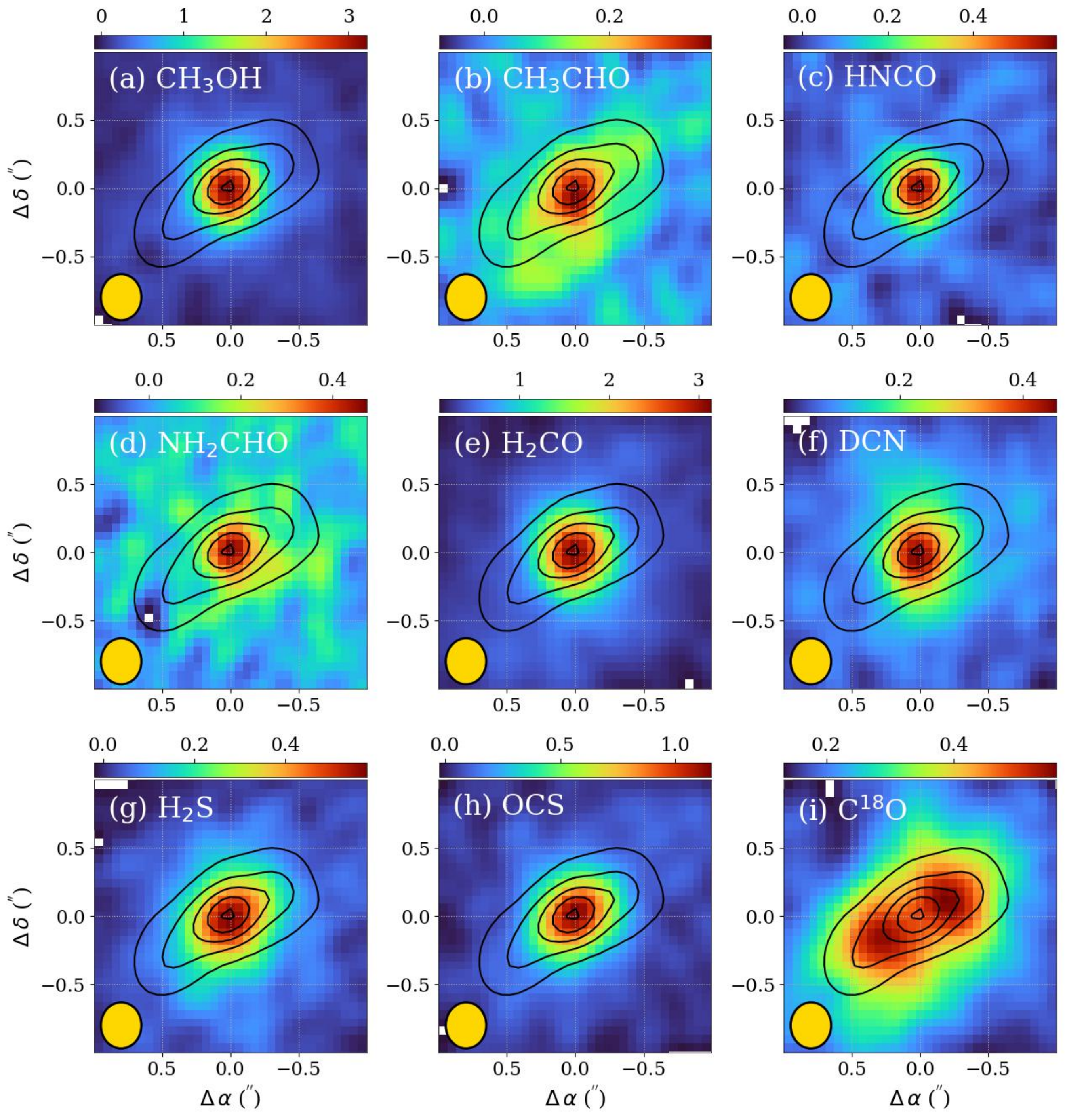}
\caption{\label{fig:img_appx} 
The integrated intensity images of selected molecules (rainbow rasters) overlaid with 1.3~mm continuum (black contours). 
The 1.3~mm continuum contour levels start from 10$\sigma$ with a step of 20$\sigma$. 
The integration ranges are 7.5~\kmpers\ for CH$_3$OH (a), H$_2$CO (e), DCN (f), H$_2$S (g), and OCS (h), 5~\kmpers\ for CH$_3$CHO (b), HNCO (c), and NH$_2$CHO (d), and 4.05~\kmpers\ for C$^{18}$O (f) with a center velocity at $v_\mathrm{LSR}=10$~\kmpers. 
}
\end{figure*}

Table \ref{tab:molec} shows the transition information in this report. 
Figure \ref{fig:img_appx} displays the integrated intensity images of the molecular species. 

\begin{deluxetable}{llllllc}
\label{tab:molec}
\caption{Information of the Transitions}
\tablehead{
\colhead{Formula} & \colhead{$f_\mathrm{rest}$} & \colhead{$E_\mathrm{u}/k_\mathrm{B}$} & \colhead{$g_\mathrm{u}$} & \colhead{$\log_{10}(A_\mathrm{ij}/\mathrm{s}^{-1})$} & \colhead{Quantum Numbers} & \colhead{Database} \\
\colhead{} & \colhead{(MHz)} & \colhead{(K)} & \colhead{} & \colhead{} & \colhead{} & \colhead{} 
}
\startdata
CO & 230538 & 17 & 5 & -6.1605 & J=2-1 & CDMS \\
C$^{18}$O & 219560 & 16 & 5 & -6.2210 & J=2-1 & CDMS \\
CH$_3$CHO & 216630 & 65 & 46 & -3.4503 & J=11-10; Ka=1; Kc=10-9; rovibSym=A & JPL \\
CH$_3$CHO & 216582 & 65 & 46 & -3.4503 & J=11-10; Ka=1; Kc=10-9; rovibSym=E & JPL \\
CH$_3$CHO & 231484 & 108 & 50 & -3.4093 & J=12-11; Ka=4; Kc=8-7; rovibSym=E & JPL \\
CH$_3$CHO & 231457 & 108 & 50 & -3.4093 & J=12-11; Ka=4; Kc=9-8; rovibSym=A & JPL \\
CH$_3$OH & 218440 & 45 & 36 & -4.3292 & J=4-3; Ka=2-1; Kc=3-2; rovibSym=E & CDMS \\
CH$_3$OH & 216946 & 56 & 44 & -4.9160 & J=5-4; Ka=1-2; Kc=4-3; rovibSym=E & CDMS \\
CH$_3$OH & 231281 & 165 & 84 & -4.7372 & J=10-9; Ka=2-3; Kc=9-6; rovibSym=A2 & CDMS \\
CH$_3$OH & 232419 & 165 & 84 & -4.7287 & J=10-9; Ka=2-3; Kc=8-7; rovibSym=A1 & CDMS \\
CH$_3$OH & 232945 & 190 & 84 & -4.6723 & J=10-11; Ka=3-2; Kc=7-9; rovibSym=E & CDMS \\
DCN & 217239 & 21 & 21 & -3.3396 & J=3-2; l2=0; kronigParity=e & CDMS \\
H$_2$CO & 218222 & 21 & 7 & -3.5501 & J=3-2; Ka=0; Kc=3-2 & CDMS \\
H$_2$CO & 218476 & 68 & 7 & -3.8037 & J=3-2; Ka=2; Kc=2-1 & CDMS \\
H$_2$CO & 218760 & 68 & 7 & -3.8020 & J=3-2; Ka=2; Kc=1-0 & CDMS \\
H$_2$CO & 216569 & 174 & 57 & -5.1413 & J=9; Ka=1; Kc=8-9 & CDMS \\
H$_2$S & 216710 & 84 & 5 & -4.3123 & J=2; Ka=2-1; Kc=0-1 & CDMS \\
HNCO & 219798 & 58 & 21 & -3.8329 & J=10-9; Ka=0; Kc=10-9 & CDMS \\
HNCO & 218981 & 101 & 21 & -3.8470 & J=10-9; Ka=1; Kc=10-9 & CDMS \\
NH$_2$CHO & 233897 & 94 & 23 & -3.0645 & J=11-10; Ka=3; Kc=9-8 & CDMS \\
NH$_2$CHO & 233735 & 115 & 23 & -3.0933 & J=11-10; Ka=4; Kc=8-7 & CDMS \\
NH$_2$CHO & 233746 & 115 & 23 & -3.0933 & J=11-10; Ka=4; Kc=7-6 & CDMS \\
NH$_2$CHO & 233594 & 142 & 23 & -3.1330 & J=11-10; Ka=5; Kc=6-5 & CDMS \\
NH$_2$CHO & 233594 & 142 & 23 & -3.1330 & J=11-10; Ka=5; Kc=7-6 & CDMS \\
NH$_2$CHO & 233528 & 174 & 23 & -3.1863 & J=11-10; Ka=6; Kc=5-4 & CDMS \\
NH$_2$CHO & 233528 & 174 & 23 & -3.1863 & J=11-10; Ka=6; Kc=6-5 & CDMS \\
OCS & 218903 & 100 & 37 & -4.5174 & J=18-17; & CDMS \\
OCS & 231061 & 111 & 39 & -4.4463 & J=19-18; & CDMS
\enddata
\end{deluxetable}

\section{YSO Modeling \label{appx:YSO}}
\resetapptablenumbers

The photometric data as well as the corresponding references are shown in Table \ref{tab:SED}. 

\begin{deluxetable}{lrrll}[tbp!]
\label{tab:SED}
\caption{SED Datapoint}
\tablehead{
\colhead{} & \colhead{$\lambda$} & \colhead{$F_\nu$} & \colhead{$\sigma(F_\nu)$} \\
\colhead{} & \colhead{(\micron)} & \colhead{(Jy)} & \colhead{(Jy)} 
}
\startdata
 WISE  & 3.4 & 2.02e-04 & 1.23e-05 \\
   & 4.6 & 3.95e-03 & 9.43e-05 \\
   & 22 & 1.85e-01 & 4.41e-03 \\
 \hline
 AKARI & 65 & 1.23e+01 & 5.93e-01 \\
   & 140 & 4.32e+01 & 2.25e+00 \\
   & 160 & 4.00e+01 & 8.33e+00 \\
 \hline
 SCUBA  & 850 & 1.91e+00 & 1.42e+00 \\
\enddata
\tablecomments{
The SED flux data points were collected from \citet{2020Dutta_ALMASOP}.
}
\tablerefs{
WISE: \citet{2010Wright_SED_WISE}
AKARI PSC: \citet{2010Ishihara_SED_AKARI_IRC};
JCMT/SCUBA: \citet{2008Francesco_JCMTS_submm}; 
}
\end{deluxetable}


\bibliography{REFERENCE.bib}{}
\bibliographystyle{aasjournal}




\end{CJK*}
\end{document}